\documentclass[floatfix, reprint, aps, pra, amsmath, amssymb]{revtex4-2}

\usepackage{graphicx}
\usepackage{dcolumn}
\usepackage{bm}
\usepackage{verbatim}
\usepackage{braket}
\usepackage{enumitem} 
\usepackage{soul}
\usepackage{textcomp, gensymb} 
\usepackage{hyperref}
\usepackage{makecell}
\usepackage{xcolor}
\usepackage{float}
\usepackage[font=small]{caption}
\usepackage[labelformat=empty, position=top]{subcaption}
\usepackage[export]{adjustbox}

\begin{document}

\preprint{APS/123-QED}

\title{Theoretical study of the ECRIPAC accelerator concept}

\author{Andrea Cernuschi}
 \email{andrea.cernuschi@lpsc.in2p3.fr}
\author{Thomas Thuillier}%
 \email{thomas.thuillier@lpsc.in2p3.fr}
\affiliation{%
 Université Grenoble Alpes, CNRS, Grenoble INP, LPSC-IN2P3, 38000 Grenoble, France
}%
\author{Laurent Garrigues}
\email{laurent.garrigues@laplace.univ-tlse.fr}
\affiliation{Université de Toulouse, Toulouse INP, CNRS, LAPLACE, Toulouse, France}


\begin{abstract}

The Electron Cyclotron Resonance Ion Plasma ACcelerator (ECRIPAC) is an original concept for a plasma-based particle accelerator able to generate pulsed ion beams with adjustable energy, targeting mostly medical applications.
This paper thoroughly reviews the working principle and physical theory behind the ECRIPAC accelerator concept, incorporating significant corrections to the existing limited literature on the subject, making it a suitable reference for future studies. Mathematical derivations for several physical formulas are also included.
Moreover, a detailed theoretical investigation of the stability condition for the ion acceleration is presented, highlighting more stringent limitations than previously anticipated. Next, the impact of several physical parameters on the accelerator design is analyzed, providing an overview of achievable external fields and plasma characteristics allowing a stable ion acceleration.
\end{abstract}
\maketitle


\section{Introduction}
\label{sec:intro}

The Electron Cyclotron Resonance Ion Plasma ACcelerator (ECRIPAC)~\cite{Geller-ecripac} is an original accelerator concept developed by R. Geller and K. Golovanivsky, able to produce very energetic pulsed ion beams, theoretically up to hundreds of MeV per nucleon. The main advantages of this device are its compact dimensions, due to its nature as a plasma-based accelerator, the simple and well mastered technology required for its operation and the possibility to adjust the ions' energy.
In its conception, ECRIPAC applications were envisioned mainly in the medical field~\cite{Schwartz-ecripac,Ishibashi-ecripac,Inoue-ecripac}, but it can be of great use also for other disciplines, such as nuclear physics and surface treatments.
The main idea behind ECRIPAC lies in the superposition of two known and experimentally verified physical principles: gyromagnetic autoresonance from gyro-resonant accelerators (GYRAC)~\cite{Golovanivsky-autoresonant_acceleration,Golovanivsky-gyromagnetic_autoresonance,Golovanivsky-gyrac} and ion entrainment inside magnetic field gradients from the so-called PLEIADE accelerators~\cite{Consoli-plasma_acceleration,Bardet-ion_entrainment, Bardet-pleiade}.
\\
Gyromagnetic autoresonance (GA)~\cite{Golovanivsky-autoresonant_acceleration,Golovanivsky-gyromagnetic_autoresonance,Golovanivsky-gyrac} is a physical mechanism able to increase the energy of an electron population confined in a time-increasing magnetic field by interaction with an injected microwave radiation, whose frequency satisfies the electron cyclotron resonance (ECR) condition.
In particular, GA exploits the strong nonlinearity of ECR, arising from the mass variation of particles moving at relativistic velocities. This latter behavior can be expressed as

\begin{equation}
    \Omega(t)=\frac{eB}{m\gamma(t)}=\frac{\Omega_0}{\gamma(t)}\;,
    \label{eq:cyclotron_frequency_shift}
\end{equation}

where $\Omega$ is the cyclotron angular frequency of an electron immersed in a constant magnetic field with intensity \textit{B}, $\gamma$ is the Lorentz factor of the electron, \textit{e} is the fundamental charge, \textit{m} is the electron mass, and \textit{t} is the time. The subscript \textit{0} is used to indicate a quantity for a non-relativistic particle.
\\
According to Eq.~\ref{eq:cyclotron_frequency_shift}, the synchronism between an injected high-frequency (HF) wave and the particle gyro-motion can be maintained through an external magnetic field growing in time $B(t)$.
If the magnetic field satisfies some conditions related to a sufficiently smooth time behavior, the electrons are accelerated in a regime characterized by phase trapping. This mechanism leads to oscillations of the phase between the electron velocity vector and the injected HF wave around a favorable regime of acceleration independently of its initial value, allowing the electron energy to increase quasi-synchronously with the magnetic field growth in a so-called gyromagnetic autoresonance regime. Hence, the electron energy rises as

\begin{equation}
    \gamma (t) \approx \frac{B(t)}{B_{0}}\;,
    \label{eq:gyrac_autoresonance}
\end{equation}

where $B_{0}$ is the magnetic field satisfying Eq.~\ref{eq:cyclotron_frequency_shift} for a non-relativistic electron and an injected HF wave with frequency $\Omega_0$.
In the standard design of GYRAC accelerators, the charged particles are trapped in a magnetic mirror trap, and the electron acceleration leads to an anisotropic electron velocity distribution in the direction perpendicular to the magnetic field lines.
\\
Ion entrainment~\cite{Consoli-plasma_acceleration,Bardet-ion_entrainment, Bardet-pleiade} takes place when the electron population of a plasma is accelerated due to a magnetic field gradient, displacing it from the ions. The local difference in ion and electron densities generates a space-charge field, which slows down the electrons and accelerates the ions, leading to a net motion of the latter in the direction of the force acting on the electrons.
Inside ECRIPAC, a high-energy electron population with a strongly anisotropic energy distribution in the perpendicular direction is immersed in a magnetic field with a negative spatial gradient along the accelerator axis. This leads to a gain in the parallel component of the electron energy at the expense of the perpendicular one due to $\nabla$B force, which in turn onsets the ion entrainment.
It has been experimentally observed that this phenomenon determines similar final axial velocity for the electrons and ions~\cite{Bardet-ion_entrainment}.
Thus, the overall effect is a net conversion of perpendicular electron energy ($W_{e\perp}$) in parallel ion energy ($W_{i\parallel}$)

\begin{equation}
    W_{e\perp}^{in}-W_{e\perp}^{fin}\approx W_{i\parallel}^{fin} \;,
    \label{eq:energy_balance_pleiade}
\end{equation}

where the index \textit{in} and \textit{fin} refer respectively to the initial and final energies. The orientations of the parallel and perpendicular directions are considered with respect to the magnetic field lines inside the system.
\\
The combination of these two phenomena allows building a device that should be theoretically able to accelerate a low pressure plasma by using well established and commercially available technologies, consisting of static and pulsed coils to generate the overall magnetic field structure and of a microwave injector.
\\
Despite the great potential of the device, ECRIPAC has almost never been studied after its initial conception, except for minor variations in the design of the magnetic field system~\cite{Geller-design_ecripac,Ishibashi-ecripac,Inoue-ecripac}.
Moreover, no prototypes have been realized to date. This turn of events can probably be attributed to an important calculation mistake in the minimum electron energy required for the PLEIADE phase stability (Eq.~\ref{eq:ecripac_gamma_min}) reported in the initial paper~\cite{Geller-ecripac}, greatly overestimating the accelerator capability.
This error has been corrected in the following publications~\cite{Geller-design_ecripac}, but still presents a discrepancy by a factor of 2 with respect to the results from some GANIL internal reports~\cite{Bertrand-report1, Bertrand-report2}. This difference arises from a different hypothesis in the estimation of the self-consistent electric field. Since the GANIL documents adopt a more reasonable approximation, lead to more realistic results, and include a more complete treatment of all phases of the ECRIPAC working cycle, their approach has been adopted as the basis for the theoretical treatment and calculations presented in this paper. All the theory has been carefully reviewed and corrected where necessary.
It is worth noting that these internal documents have never been published, hence the most recent literature on ECRIPAC~\cite{Ishibashi-ecripac,Inoue-ecripac} is based on incomplete and possibly wrong premises.
\\
This work presents an in-depth theoretical study based on the current state of the ECRIPAC physics, to better comprehend the limitations of the accelerators and the requirements for the different sections of the device.
First, the physics principles of the ECRIPAC accelerator concept are presented in Section~\ref{sec:wp}. Further details on the ECRIPAC theory are proposed in Appendix~\ref{app:comp} and \ref{app:pl}. Then, the design limitations associated with this device are discussed (Section~\ref{sec:des_lim}), followed by an analysis of the main parameters characterizing ECRIPAC (Section~\ref{sec:par_anal}). Finally, some conclusions and perspectives for the future work are provided (Section~\ref{sec:conclusion}).


\section{ECRIPAC working principle}
\label{sec:wp}

\begin{figure}[b]
    \centering
    \includegraphics[width=1.\linewidth, valign=t]{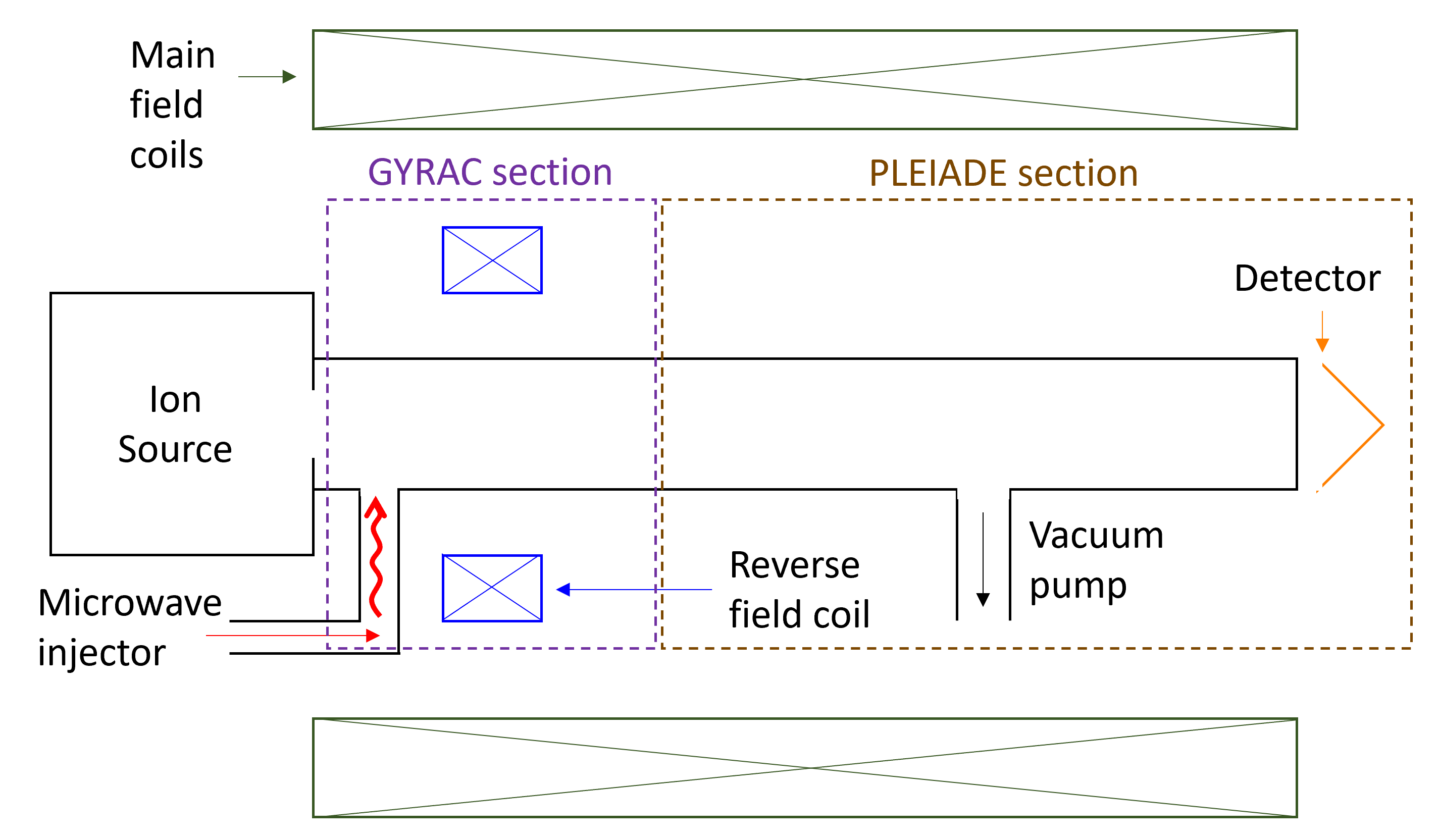}
    \caption{Schematic representation of ECRIPAC.}
    \label{fig:ecripac}
\end{figure}

\begin{figure}[b]
    \centering
    \includegraphics[width=1.\linewidth, valign=t]{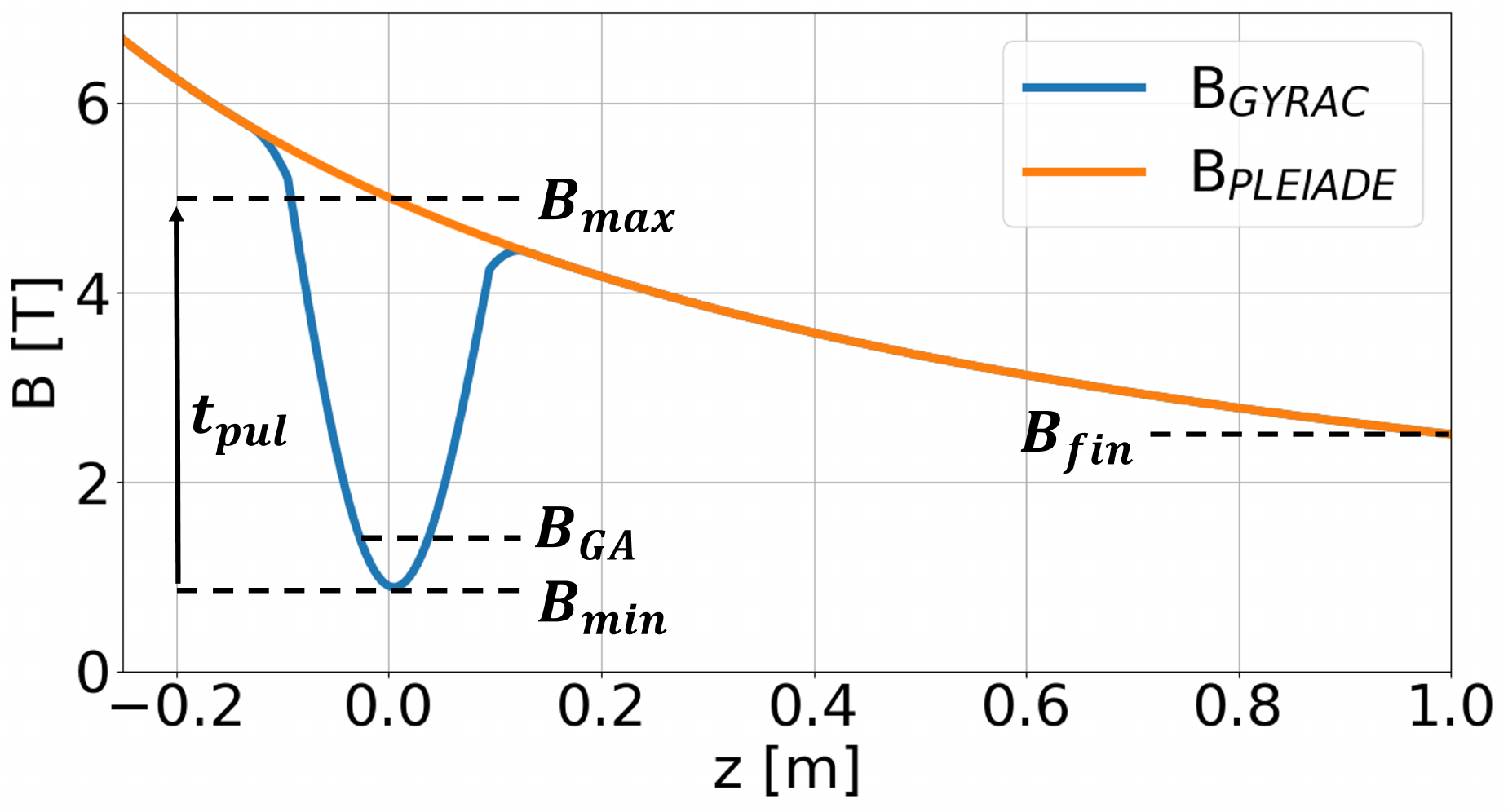}
    \caption{Schematic representation of the ECRIPAC magnetic field. $B_{min}$, $B_{GA}$ and $B_{max}$ are the magnetic field strengths at the center of the magnetic mirror at the beginning, end of the GA phase and at the end of the transient field behavior respectively, $t_{pul}$ is the time required to restore the main field, and $B_{fin}$ is the magnetic field value at the end of the PLEIADE cavity.}
    \label{fig:ecripac_Bfield}
\end{figure}

An ECRIPAC device~\cite{Geller-ecripac} is composed of three sections, each playing a role in the overall acceleration process (see Fig.~\ref{fig:ecripac}):

\begin{itemize}
    \item An ion source, used as an injector of low pressure highly ionized plasma in free diffusion through differential pumping. An ECR ion source~\cite{Geller_book} has been proposed as the plasma injector due to its ability to generate a highly ionized plasma.
   
    \item A GYRAC section, which includes a reverse field pulsed coil, a metallic resonant cavity, and a microwave injection port. The function of this section is to heat the electron population through gyromagnetic autoresonance and compress it in the shape of a thin disk.
   
    \item A PLEIADE section, constituted by a metallic cavity and an axial distribution of coils, generating a long magnetic gradient encompassing the GYRAC section. Its function is to accelerate the ions to very high energy.
\end{itemize}

\begin{figure}[t]
    \centering
    \includegraphics[width=1.\linewidth, valign=t]{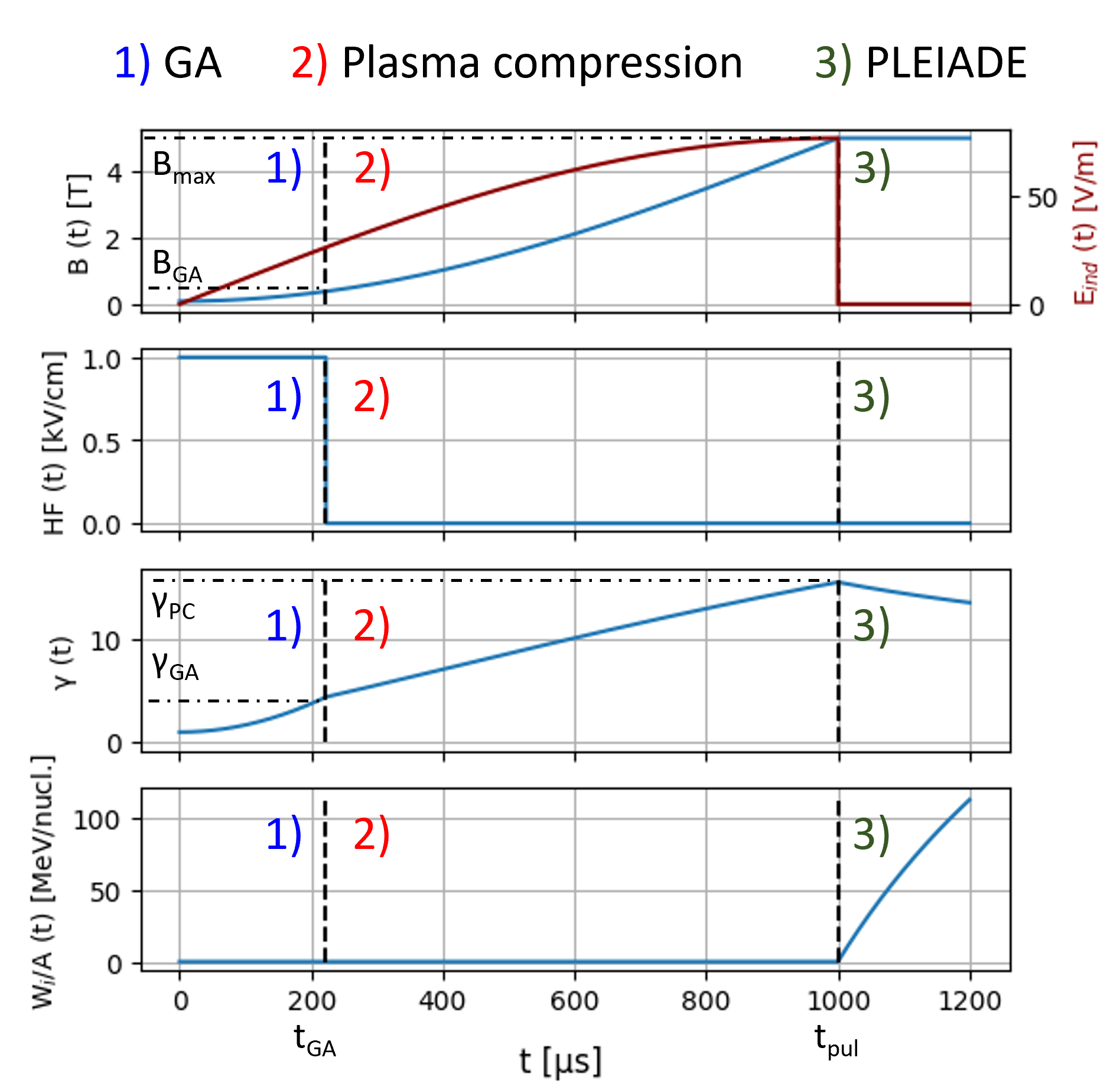}
    \caption{Time evolution of the magnetic field $B (t)$, electric field induced by $B (t)$ ($E_{ind} (t)$), HF wave intensity $\textit{\mbox{HF}}(t)$, electron normalized energy $\gamma (t)$ and ion kinetic energy per nucleon $W_i/A (t)$ during the three successive phases of ECRIPAC (GA, Plasma compression and PLEIADE). The indexes \textit{GA} and \textit{PC} indicate the value of the corresponding quantity at the end of the GA and plasma compression phase respectively. $B_{max}$ and $t_{pul}$ have been defined in Fig.~\ref{fig:ecripac_Bfield}. All quantities are evaluated in the location of the minimum intensity of the magnetic mirror, on the coil axis, except for $\gamma (t)$ and $W_i/A (t)$ during the PLEIADE phase and $E_{ind} (t)$, evaluated at a radius of 2 cm.}
    \label{fig:chronograph}
\end{figure}

A typical ECRIPAC axial magnetic field profile is reported in Fig.~\ref{fig:ecripac_Bfield}. The main static magnetic field (PLEIADE section) is displayed in orange. The maximum effect of the reversed pulsed magnetic field (GYRAC section) is superimposed in blue on the magnetic field profile. The parameters appearing in this figure are detailed later.
Fig.~\ref{fig:chronograph} presents the temporal evolution of five key parameters of the ECRIPAC acceleration. Namely, the time evolution of the pulsed magnetic mirror well $B (t)$, the induced electric field $E_{ind} (t)$, the HF wave intensity $\textit{\mbox{HF}}(t)$, the normalized electron energy $\gamma(t)$ and the ion kinetic energy per nucleon $W_i/A(t)$.
Helped with these two figures, the working principle of the ECRIPAC acceleration is detailed in the following subsections.

\subsection{Gyromagnetic autoresonance (GA)}
\label{sec:wp-ga}

The ECRIPAC process starts by triggering the reverse pulsed coil, delivering an axial sinusoidal magnetic field intensity with a quarter period noted $t_{pul}$.
When  the pulsed coil reaches its maximum field (at $t=0$ in Fig.~\ref{fig:chronograph}), an axial magnetic mirror is created around $z=0$ (see Fig.~\ref{fig:ecripac_Bfield}) and the cavity is filled with the plasma exiting the source.
At this time, the maximum peak field outside the magnetic mirror well is approximately equal to $B_{max}$, while the minimum field (noted $B(0)=B_{min}$) is such that
 
\begin{equation}
    B_{min}\lesssim B_{ECR} = \frac{m}{e} \omega_{HF},
\end{equation}

where $B_{ECR}$ is the ECR magnetic field for thermal electrons, and $\omega_{HF}$ is the input high frequency (HF) angular frequency. At $t=0$, the HF power is injected into the cavity. The electrons in the plasma are immediately heated through the ECR mechanism and, since $B(t)$ increases with time, a gyromagnetic autoresonance (later referred to as GA) occurs ~\cite{Geller-ecripac}.
The presence of a magnetic mirror is necessary to confine and heat the electrons arriving from the injector.
\\
If the reverse pulsed magnetic field temporal evolution respects the GA hypothesis for a sufficiently slow-varying field in time~\cite{Golovanivsky-gyromagnetic_autoresonance}, the energy of an electron inside ECRIPAC should on average follow Eq.~\ref{eq:gyrac_autoresonance}, yielding a final value at the end of the GA phase independent of the electric field intensity and equal to

\begin{equation}
    W_{\perp e,GA}\approx\gamma_{GA}m_e c^2\approx0.51\left(\frac{B\left(t_{GA}\right)}{B_{min}}-1\right) \;MeV\;,
    \label{eq:ecripac_gyrac_energy}
\end{equation}

where the subscript \textit{GA} refers to a quantity at the end of the GA phase.


\subsection{Plasma compression (PC)}
\label{sec:wp-comp}

The plasma compression phase~\cite{Geller-ecripac} starts as soon as the injection of the HF wave inside the resonant cavity is stopped (at $t=t_{GA}$), while the total magnetic mirror well $B(t)$ continues to increase up to the complete restoration of the initial main magnetic field (orange line in Fig.~\ref{fig:ecripac_Bfield}) at $t=t_{pul}$.
During this period, the energy of the electron population increases by means of the electric field induced by the time variation of the magnetic field (see $E_{ind} (t)$ evolution in Fig.~\ref{fig:chronograph}). However, the most important phenomenon taking place during this phase is the compression of the plasma, leading to increased electron and ion densities, which are crucial factors for an appropriate ion acceleration taking place afterward. Indeed, under the hypothesis of slowly varying magnetic field, it is possible to demonstrate that during this phase there exist the following adiabatic constants of motion in the dynamics of a single electron propagating inside the ECRIPAC magnetic field configuration (see Appendix~\ref{app:comp} for details on derivation)

\begin{equation}
    \frac{p_e^2}{B}=const \hspace{0.5 cm} r_{Lar}^2B=const \hspace{0.5 cm} r_{ec}^2B=const\; ,
    \label{eq:ecripac_compression_constants}
\end{equation}

where $p_e$ is the electron momentum, while $r_{Lar}$ and $r_{ec}$ are respectively the Larmor radius of the electron orbit and the distance between the orbit center and the cavity axis.
The increasing magnetic field over time leads to a decrease of both $r_{Lar}$ and $r_{ec}$, thus radially compressing the plasma. The general configuration of the plasma is conserved during this process, keeping the initial disk shape unaltered.
Moreover, it is possible to demonstrate that the axial oscillatory motion of the electrons inside the magnetic mirror is damped in time, leading also to an axial compression of the plasma disk (see Appendix~\ref{app:comp} for details).
The plasma’s tendency to maintain neutrality also induces a motion of the ions, which slowly follow the electrons along their trajectories, leading to a global compressive motion of the plasma.
\\
Finally, starting from the equation of motion for the electrons, it is possible to obtain the time evolution of the particle cyclotron angular frequency, radial distance of the particle from the cavity center and Lorentz factor during the plasma compression phase:

\begin{eqnarray}
    \gamma(t)=\sqrt{1+(\gamma_{GA}^2-1)\frac{B(t)}{B_{GA}}} \;,
    \label{eq:ecripac_compression_gamma}
    \\
    r(t)=r_{GA}\sqrt{\frac{B_{GA}}{B(t)}} \;,
    \label{eq:ecripac_compression_r}
    \\
    \Omega(t)=\omega_{HF}\gamma_{GA}\frac{B(t)}{B_{GA}}\frac{1}{\sqrt{1+(\gamma_{GA}^2-1)\frac{B(t)}{B_{GA}}}} \;.
    \label{eq:ecripac_compression_omega}
\end{eqnarray}

Further details regarding the derivation of these results can be found in Appendix~\ref{app:comp}.
For example, let's consider an accelerator using a heating frequency $f_{HF}=2.45$ GHz and characterized by a magnetic field that increases from 0.0875 T up to 5 T. The estimated values of the electron Lorentz factor, cyclotron angular frequency and radial distance from the cavity center at the end of both the GA and plasma compression phases are reported in Table~\ref{tab:estimations}. It can be clearly seen the fundamental role of the plasma compression phase, strongly compressing the electron cloud and increasing its energy.

\begin{table}[b]
\centering
\caption{\label{tab:estimations}%
Estimation of several electron physical quantities for an accelerator using a heating frequency $f_{HF}=2.45$ GHz and characterized by a magnetic field which increases from 0.0875 T up to 5 T.
}
\begin{tabular}{l @{\hskip 10pt} c @{\hskip 10pt} c @{\hskip 10pt} c @{\hskip 10pt} c}
\hline\hline
Value at the end of & $B$ [T] & $\gamma$ & $r$ [cm] & $\Omega$ \\
\hline
GA phase & 0.5 & 5.71 & 1.92 & $\omega_{HF}$ \\
PC phase & 5 & 17.81 & 0.61 & $3.2\cdot\omega_{HF}$ \\
\hline\hline
\end{tabular}
\end{table}


\subsection{PLEIADE (PL)}
\label{sec:wp-pl}

The PLEIADE phase~\cite{Geller-ecripac} is the last part of the ECRIPAC working cycle, whose objective is the acceleration of the ion population through the ion entrainment mechanism.
Thus, the $\nabla B$ force arising from the negative gradient of the magnetic field converts the electron perpendicular velocity into parallel velocity, while the space-charge field generated inside the plasma pulls the ions and accelerates them.
An important condition that the PLEIADE magnetic field must satisfy is~\cite{Geller-design_ecripac} 

\begin{equation}
    \frac{d^2B}{dz^2}>0,
    \label{eq:ecripac_pleiade_condition}
\end{equation}

where $z$ is the axis of the cavity. In the absence of this requirement, the plasma can be subjected to macroscopic instabilities, deforming its initial shape to a long cylinder and preventing the confinement and acceleration of the electrons.
A physical phenomenon mitigating this condition is given by the diamagnetic character of the plasma itself, which generates a magnetic field oriented oppositely to the external one. This leads to the creation of a local magnetic well, which favors the requirements given by Eq.~\ref{eq:ecripac_pleiade_condition}~\cite{Geller-design_ecripac}.
\\
Starting from the electron equation of motion, it is possible to demonstrate that the constants of motion expressed in Eq.~\ref{eq:ecripac_compression_constants} are still valid for a sufficiently low magnetic field gradient (considering only the azimuthal component of the electron momentum, see Appendix~\ref{app:pl} for details on derivation).
Thus, it is necessary to pay particular attention when choosing the final minimum field for the PLEIADE phase $B_{fin}$, since an excessively low value can lead to the spatial divergence of the electron bunch, which gets deconfined at the wall.
\\
Considering a plasma composed of ions with mass $A m_a$ and electric charge $Q=Ze$ (where $m_a$ is the atomic mass unit, $A$ and $Z$ the mass and charge numbers respectively), with a number of ions much lower than electrons ($N_i \ll N_e$), the energy of the ions per nucleon at the end of the accelerating process can be estimated as (see Appendix~\ref{app:pl} for details on derivation)

\begin{equation}
    \frac{W_i}{A}\approx\frac{1}{2}\frac{\gamma_{PC}^2-1}{\gamma_{PC}^2}m_{a} c^2\left(1-\frac{B_{fin}}{B_{max}}\right) \;,
    \label{eq:ecripac-ion_energy_simple}
\end{equation}

where \textit{c} is the speed of light, while $\gamma_{PC}$ is the Lorentz factor of electrons at the end of the compression phase.
Eq.~\ref{eq:ecripac-ion_energy_simple} will be exploited and analyzed in more details in Section~\ref{sec:par_anal}.
The accelerator stability during the PLEIADE phase strongly depends on the magnetic field profile, and it is mainly constrained by two phenomena: the stability of the electron bunch and the ion shake-out (some ions are not accelerated by the electrons and detach from the plasma).
An ion is entrained by the electrons only if the Coulomb attraction overcomes the electrons' acceleration (non shake-out condition), which can also be expressed in the following form by making explicit the PLEIADE field profile (see Appendix~\ref{app:pl} for details on derivation)

\begin{equation}
    \left|\frac{\nabla B_z}{B_z}\right|\leq \frac{2e}{m_{a}c^2}\frac{Z}{A}E_{sc}\;,
    \label{eq:ecripac_nsc}
\end{equation}

where $E_{sc}$ is the space-charge electric field generated inside the plasma.
Hence, any ion with a ratio $A/Z$ not respecting condition~\ref{eq:ecripac_nsc} is not entrained by the electron bunch and is shaken out from the plasma.
A more exact approach should consider a progressive shake-out of the ion population during their acceleration in the PLEIADE field (see Appendix~\ref{app:pl} for details).
Thus, according to Eq.~\ref{eq:ecripac_nsc}, very large magnetic field gradients will accelerate only the electron population, since all the ions will be shaken out from the plasma.
\\
On the other hand, the stability of the electron bunch during the PLEIADE stage is influenced by several factors, where the main contributions are given by the characteristic time for the spatial divergence of the electron bunch due to Coulomb repulsion and its dampening due to relativistic effects. Thus, the stability condition of the electron bunch can be written as an inequality between the radial and axial forces experienced by the electrons, expliciting the PLEIADE field profile (see Appendix~\ref{app:pl} for details on derivation)

\begin{equation}
    \left|\frac{\nabla B_z}{B_z}\right|\geq \frac{2e}{m_ec^2(\gamma^3-\gamma)}E_{sc}\;.
    \label{eq:ecripac_esc}
\end{equation}

By equating Eqs.~\ref{eq:ecripac_nsc} and \ref{eq:ecripac_esc}, it is possible to determine a minimum value for the electron Lorentz factor $\gamma_{min}$ at the end of the plasma compression phase to allow the acceleration of ions with a given $A/Z$ ratio

\begin{equation}
    \gamma_{min}^3-\gamma_{min}=\frac{m_{a}}{m_e}\frac{A}{Z} \;.
    \label{eq:ecripac_gamma_min}
\end{equation}

For example, the minimum Lorentz factor needed to accelerate protons is equal to 12.25, while for $Ar^{8+}$ it increases up to 21.
These strong requirements for the electron energies at the end of the plasma compression phase are a considerable difficulty for ECRIPAC accelerators, since they require substantially higher magnetic fields to achieve them (see Eqs.~\ref{eq:gyrac_autoresonance} and \ref{eq:ecripac_compression_gamma}).
This would introduce technological complications in the development of a prototype. Another possibility is to consider a longer GA phase (since Eq.~\ref{eq:gyrac_autoresonance} leads to a larger energy gain than Eq.~\ref{eq:ecripac_compression_gamma}), which however could lead to the insurgence of plasma instabilities and a worse plasma compression.
\\
To sum up, an ECRIPAC device must be characterized by several properties to favor the acceleration of ions:

\begin{itemize}
    \item The magnetic field gradient during the PLEIADE stage must be tuned correctly, to allow a stable acceleration of the ions to the desired energy.
    \item The ratio $A/Z$ must be low, leading to the requirement of a strongly ionized plasma.
    \item The number of accelerated electrons at the end of the plasma compression phase must be very high.
    \item The electron density must be as large as possible, implying the necessity of a proper plasma compression (See Section~\ref{sec:des_lim} for more details).
\end{itemize}


\section{Stability conditions of the accelerator}
\label{sec:des_lim}

\begin{figure}[t!]
\hspace*{\fill}
\begin{subfigure}{0.48\textwidth}
  \centering
  \includegraphics[width=0.87\linewidth]{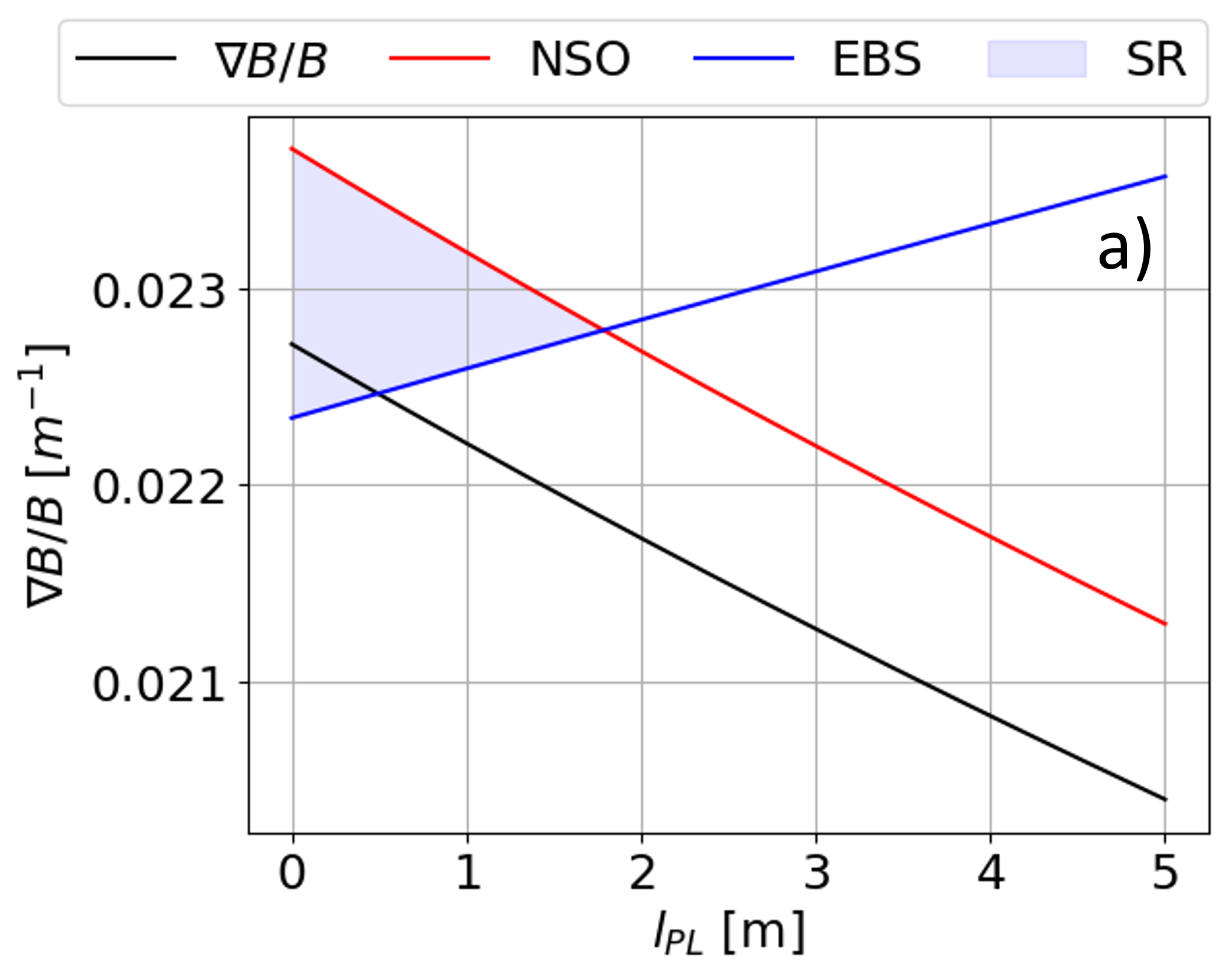}
\end{subfigure}
\hspace*{\fill}
\begin{subfigure}{0.48\textwidth}
  \centering
  \includegraphics[width=0.87\linewidth]{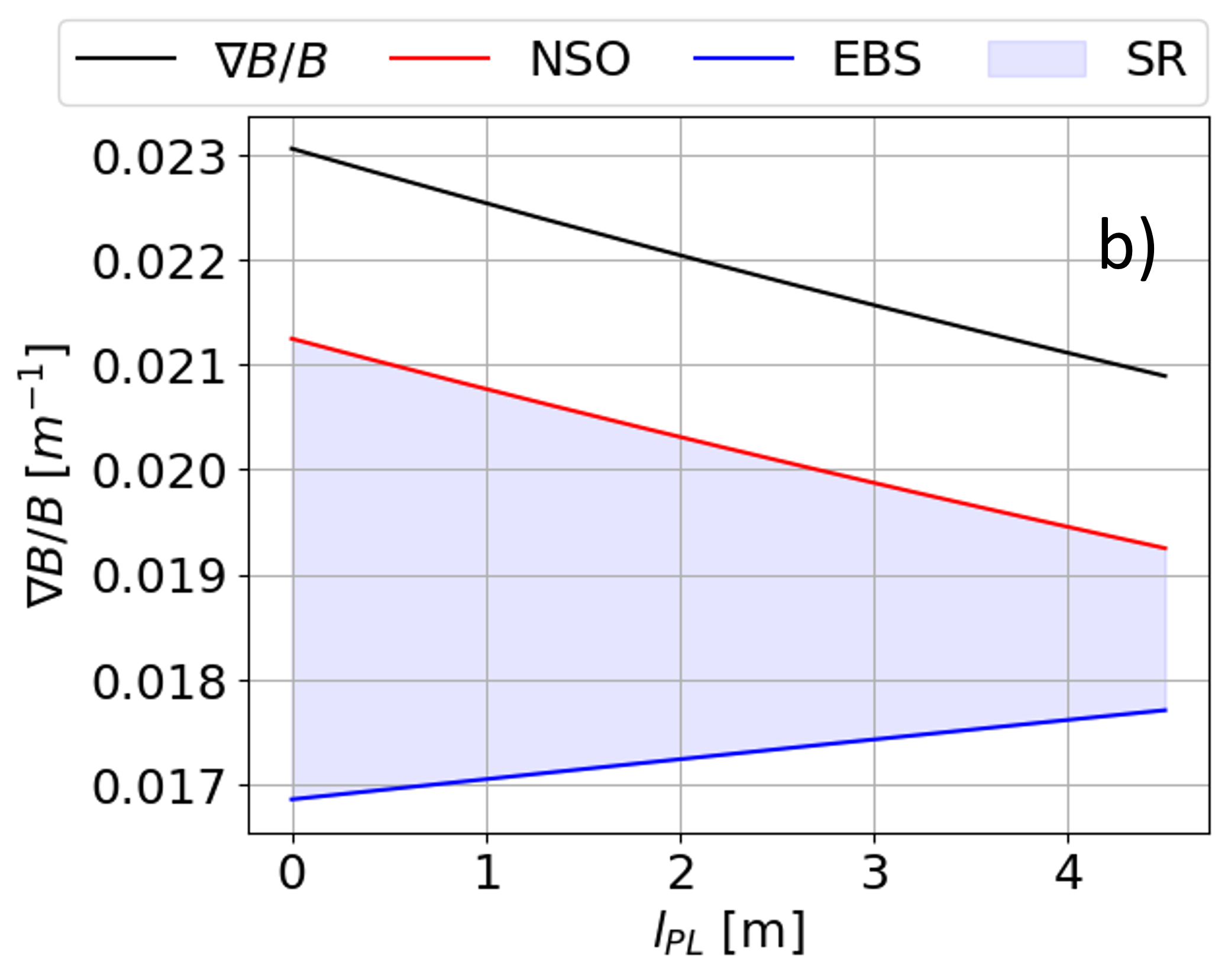}
\end{subfigure}
\hspace*{\fill}
\begin{subfigure}{0.48\textwidth}
  \centering
  \includegraphics[width=0.87\linewidth]{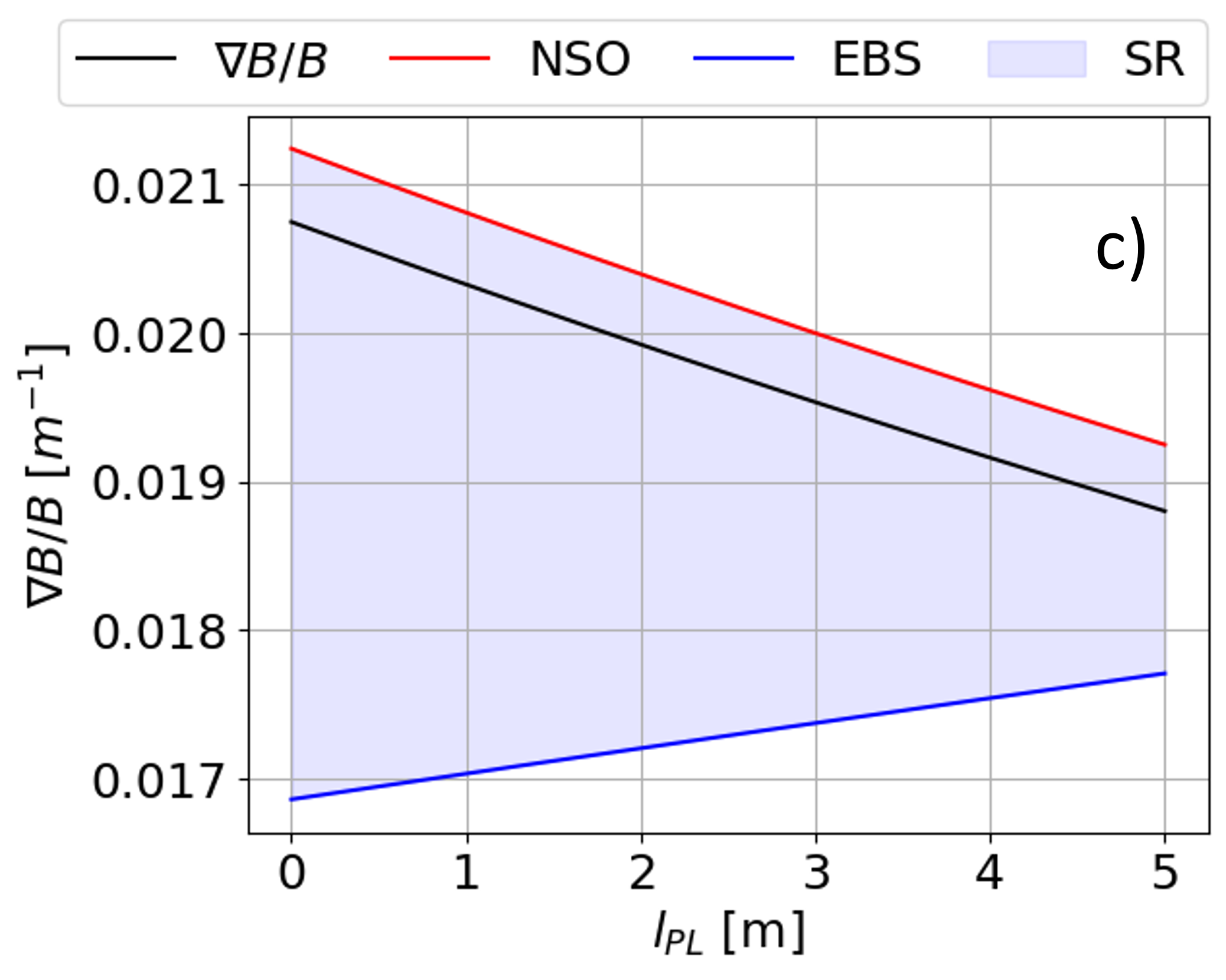}
\end{subfigure}
\caption{Representation of stability condition for $\nabla B/B$ as a function of the cavity length ($l_{PL}$) for a) $\gamma_{PC}=15.76$ and $l_{PL}=5\;m$; b) $\gamma_{PC}=16.69$ and $l_{PL}=4.5\;m$; c) $\gamma_{PC}=16.69$ and $l_{PL}=5\;m$. The ion acceleration is stable when the black line is contained inside the light blue area (SR: stability region). When the red line (NSO: non shake-out condition, upper limit for stability) lies below the blue line (EBS: electron bunch stability condition, lower limit for stability), the stability region (SR) for the accelerator does not exist. $\gamma_{PC}$ and $\gamma_{min}$ are defined in Section~\ref{sec:wp-pl}.}
\label{fig:limitations}
\end{figure}

The theoretical framework described in Section~\ref{sec:wp} still presents some shortcomings, aside from the omission of collective effects.
For this reason, this section will be devoted to a careful theoretical study of ECRIPAC, starting from the physical foundations laid down in the previous sections.
\\
As already mentioned, one of the most critical points when designing an ECRIPAC device is to ensure a stable acceleration of the ions of interest. This condition depends on the magnetic field profile for the PLEIADE section and, neglecting the progressive ion shake-out, presents both an upper and lower limit corresponding to the non shake-out condition (Eq.~\ref{eq:ecripac_nsc}) and the electron bunch stability condition (Eq.~\ref{eq:ecripac_esc}) respectively. Combining these two requirements in Eq.~\ref{eq:ecripac_gamma_min} provides a lower limit for the electron energy at the end of the plasma compression phase, which however is not sufficient to ensure a stable acceleration over the whole PLEIADE section. 
Indeed, the constants of motion valid also for the PLEIADE phase (Eq.~\ref{eq:ecripac_compression_constants}) determine a progressive increase in the size of the electron bunch along the cavity due to the negative gradient of the magnetic field. Considering the electron bunch as a thin disk and a number of ions much lower than the number of electrons ($N_i \ll N_e$), the peak value for the space-charge electric field can be approximated as the field generated by a very large and infinitesimally thin uniformly charged disk

\begin{equation}
    E_{sc}=\frac{eN_e}{2\epsilon_0 S}\propto \frac{1}{r_e^2} \; ,
    \label{eq:Ecollective}
\end{equation}

where $S$ is the surface of the disk, $r_e$ is the radius of the electron disk and $\epsilon_0$ is the vacuum permittivity constant. Thus, Eqs.~\ref{eq:ecripac_nsc} and \ref{eq:ecripac_esc} will provide progressively more stringent upper and lower limits along the accelerating cavity, narrowing or even erasing the stability region for the accelerator (Eq.~\ref{eq:ecripac_esc} becomes more stringent due to the decrease in electron energy, which is progressively converted to ion energy, see Appendix~\ref{app:pl} for details). Thus, it is necessary to have an average electron energy at the end of the plasma compression phase quite larger than the lower limit imposed by Eq.~\ref{eq:ecripac_gamma_min}, leading to a sufficient initial width for the stability region, which allows it to exist until the end of the PLEIADE section.
\\
This limitation can be better understood from the examples in Fig.~\ref{fig:limitations}, considering an accelerator design for He$^{2+}$ ($\gamma_{min}\approx15.45$): in the first case (Fig.~\ref{fig:limitations}a), since the electron energy at the beginning of the PLEIADE phase is not large enough ($\gamma_{PC}=15.76$), the accelerator loses its stability halfway through the cavity (at 1.8 m), as can be seen by the non shake-out (NSO, Eq.~\ref{eq:ecripac_nsc}) condition becoming lower than the electron bunch stability (EBS, Eq.~\ref{eq:ecripac_esc}) condition; in the other cases (Fig.~\ref{fig:limitations}b and \ref{fig:limitations}c), characterized by a larger electron energy ($\gamma_{PC}=16.69$), the stability region (SR, light blue area in Fig.~\ref{fig:limitations}) exists over the whole cavity length.
It must be emphasized that the electron energy is not the only parameter determining the existence of a stability region. Indeed, a smaller difference between the maximum and minimum value of the PLEIADE magnetic field increases the width of the stability region, at the cost of a lower final ion energy.

Moreover, the existence of a stability region does not always imply a stable ion acceleration. To ensure this latter condition, the magnetic field profile of the PLEIADE section must have a proper value of $\nabla B/B$, as can be observed in Fig.~\ref{fig:limitations}b and \ref{fig:limitations}c, namely the $\nabla B/B$ (black curve) must be strictly located in between the NSO and EBS curves (red and blue curves respectively).
It is clear that the main parameters to control $\nabla B/B$ are $B_{max}$, $B_{fin}$ and the length of the accelerating cavity ($l_{PL}=4.5$ m for Fig.~\ref{fig:limitations}b and $l_{PL}=5$ m for Fig.~\ref{fig:limitations}c), which alters the gradient of the magnetic field. $B_{max}$ and $B_{fin}$ have been considered constants in the present study, since varying these parameters often requires more stringent technological constraints.
Another important feature in this regard is obviously the shape of the PLEIADE magnetic field, which must have a negative gradient and respect Eq.~\ref{eq:ecripac_pleiade_condition}. Since the optimal profile for ion acceleration must be parallel to the non shake-out condition, the magnetic field

\begin{equation}
    B(z)=\frac{1}{B_{max}^{-1}+kz},
    \label{eq:Bpl_optimal}
\end{equation}
where
\begin{equation}
    k=\frac{B_{fin}^{-1}-B_{max}^{-1}}{l_{PL}}
    \label{eq:Bpl_optimal_2}
\end{equation}
has been adopted in all calculations to maximize the accelerator stability region.
Since many other parameters affect the ECRIPAC stability, the next part is dedicated to the investigation of the parameters' phase space resulting in a stable ECRIPAC ion acceleration.


\section{ECRIPAC parameter analysis}
\label{sec:par_anal}

\begin{figure*}[!p]
\begin{subfigure}{0.83\textwidth}
  \centering
  \includegraphics[width=1\linewidth]{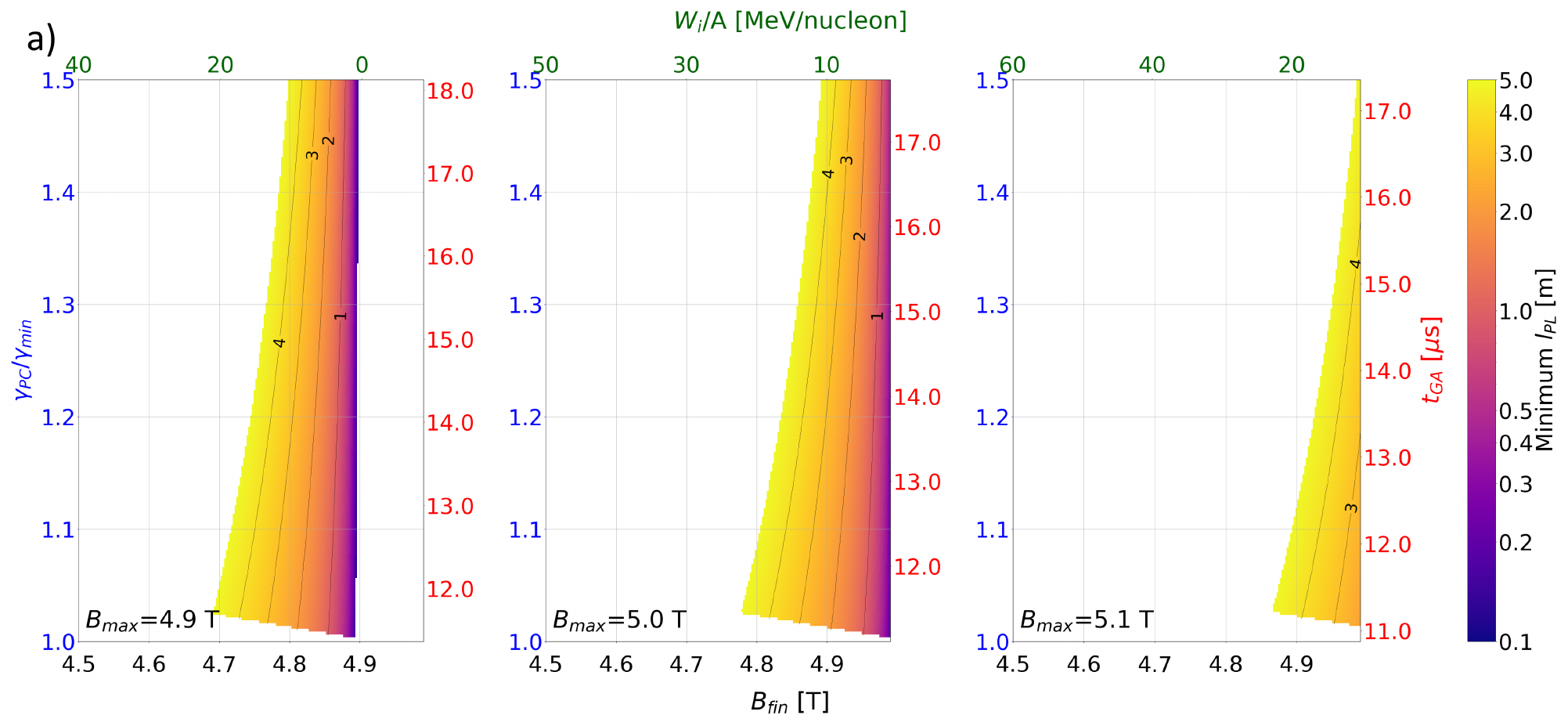}
\end{subfigure}
\begin{subfigure}{0.83\textwidth}
  \centering
  \includegraphics[width=1\linewidth]{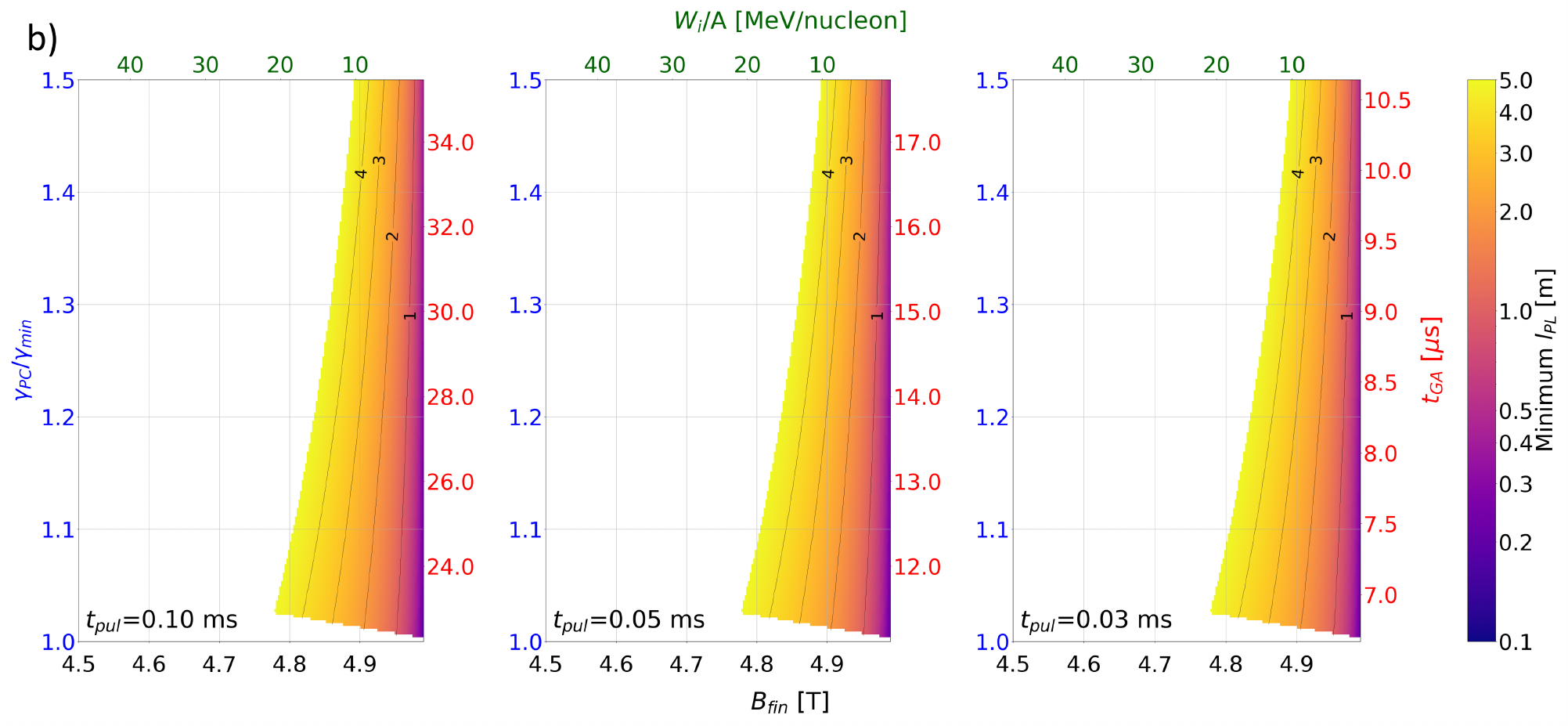}
\end{subfigure}
\begin{subfigure}{0.83\textwidth}
  \centering
  \includegraphics[width=1\linewidth]{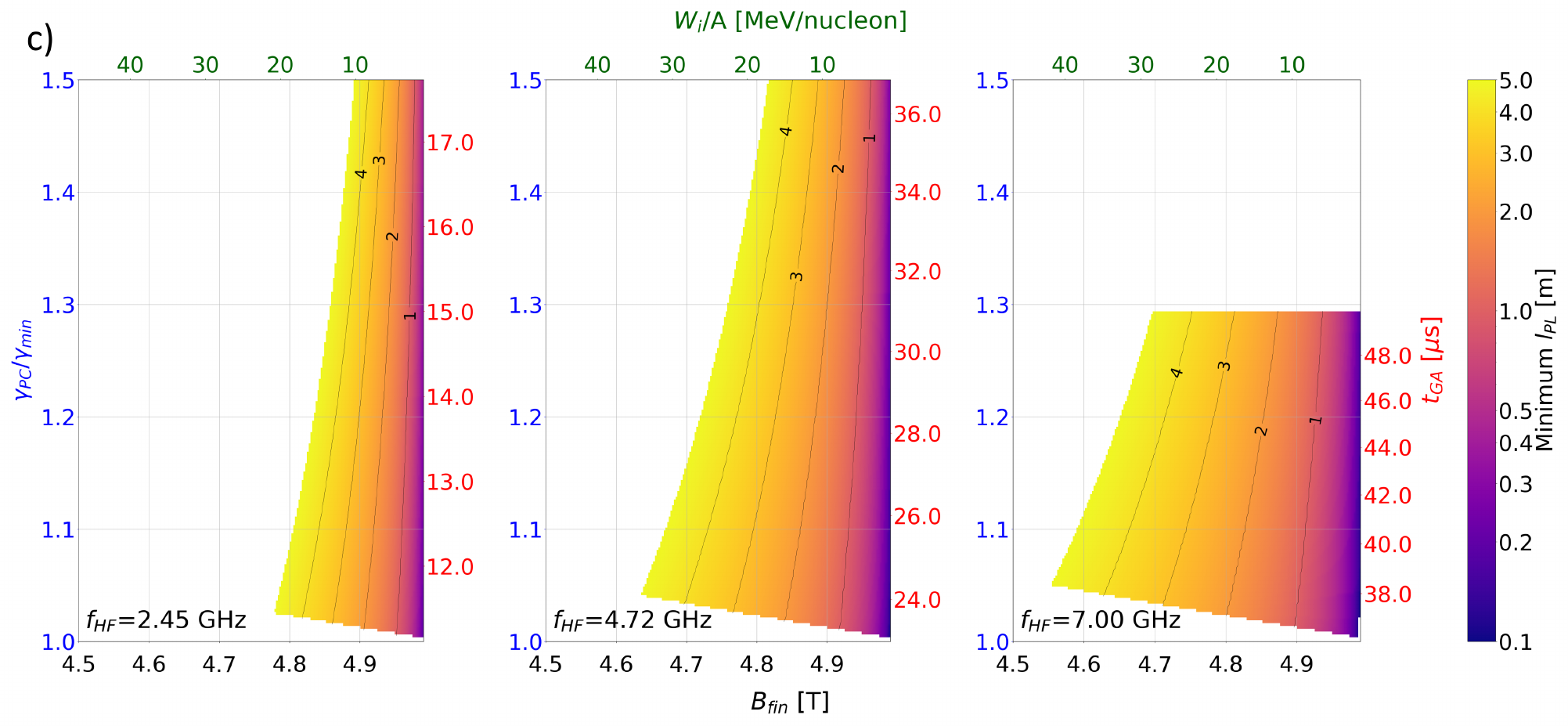}
\end{subfigure}
\caption{Stability maps of ECRIPAC for different parameters of the external fields, including a) the maximum magnetic field $B_{max}$; b) the rise time of the pulsed field $t_{pul}$; c) the heating frequency of the microwave $f_{HF}$. The nominal values of the accelerator parameters when they are not varied are $B_{max}=5$ T, $t_{pul}=50\;\mu s$, $f_{HF}=2.45$ GHz, A/Z = 2, $n_e=10 \% \; n_{cr}$ and $r_{ec,0}=2$ cm. The ion energy has been evaluated considering a negligible number of ions accelerated compared to the number of electrons (Eq.~\ref{eq:ecripac-ion_energy_simple}). The color bar representing the minimum cavity length ($l_{PL}$) required for a stable acceleration is in logarithmic scale.}
\label{fig:ext_fields}
\end{figure*}

\begin{figure*}[!p]
\begin{subfigure}{0.83\textwidth}
  \centering
  \includegraphics[width=1\linewidth]{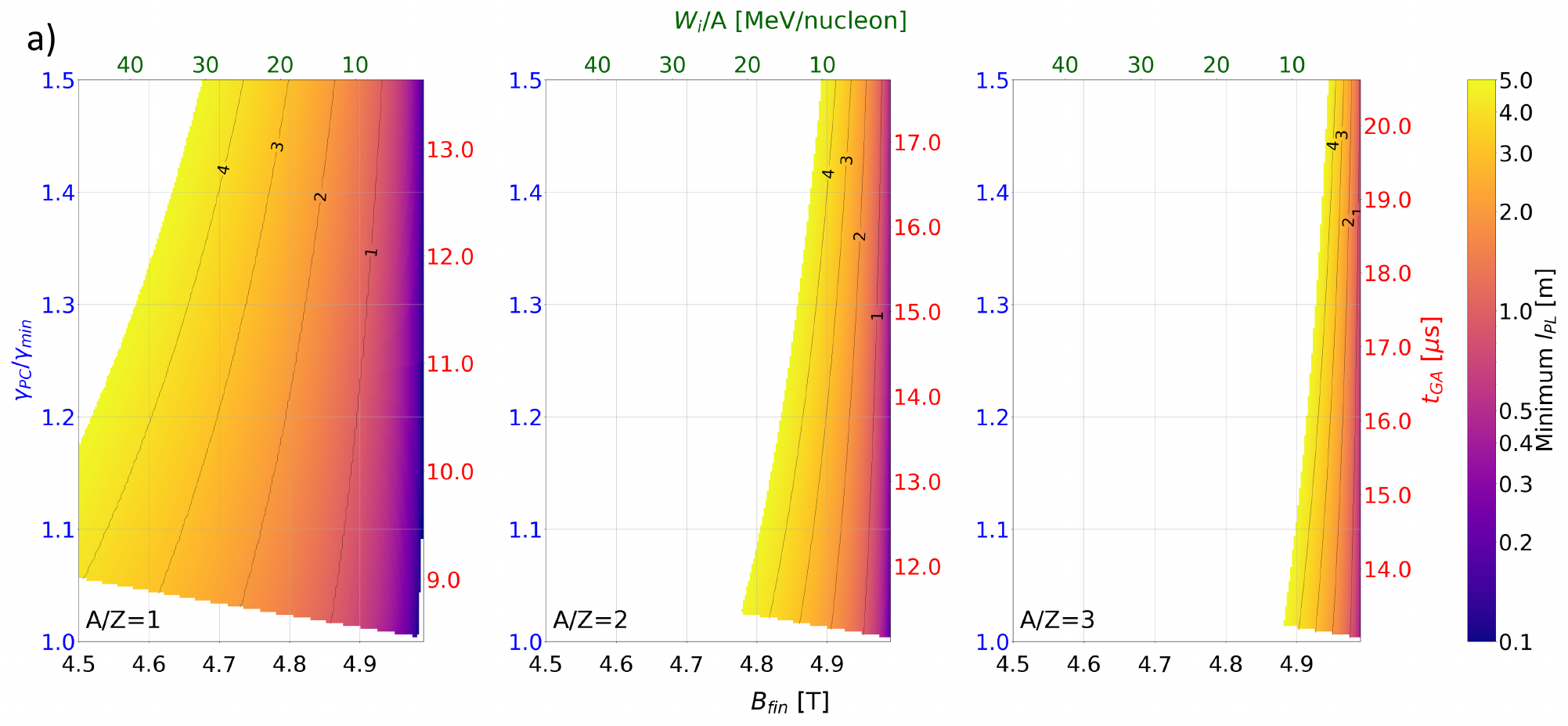}
\end{subfigure}
\begin{subfigure}{0.83\textwidth}
  \centering
  \includegraphics[width=1\linewidth]{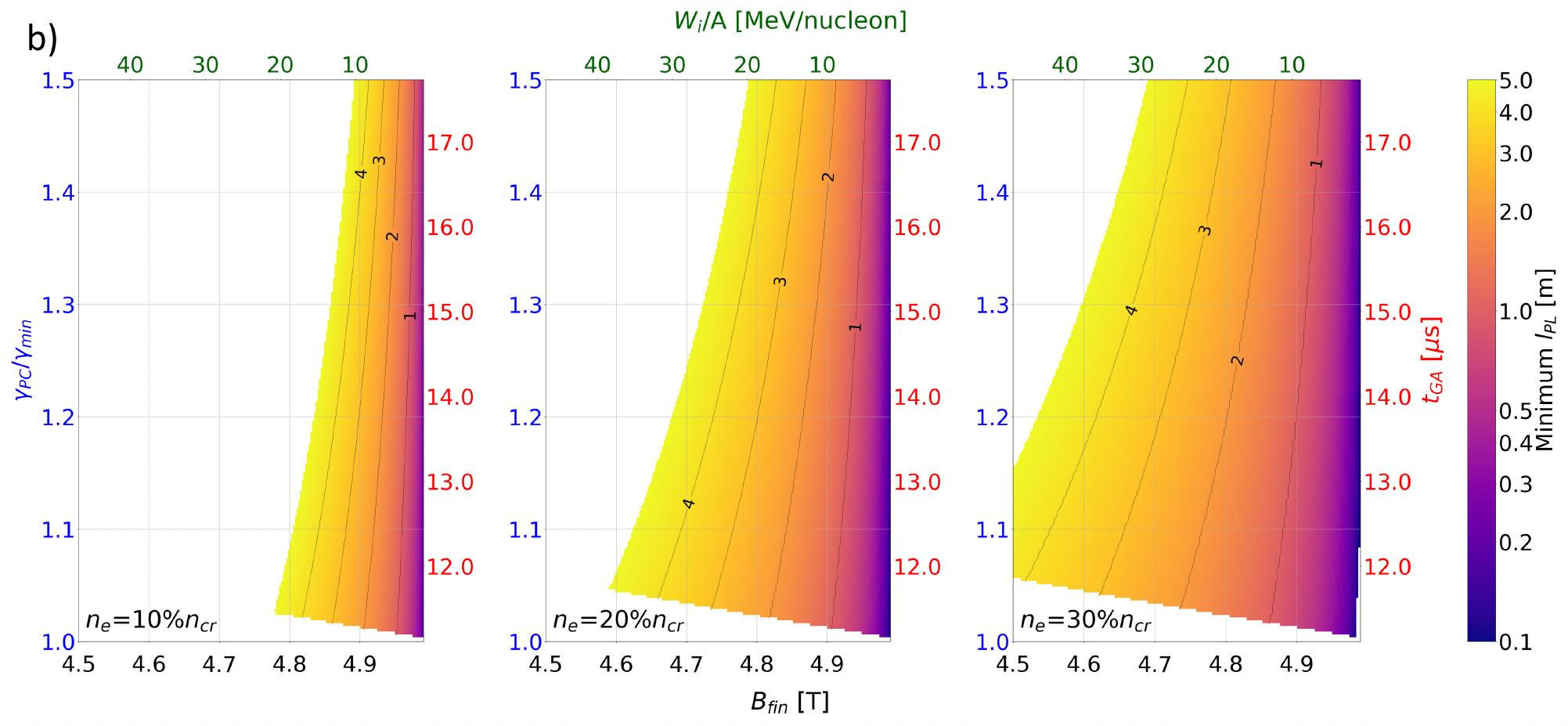}
\end{subfigure}
\begin{subfigure}{0.83\textwidth}
  \centering
  \includegraphics[width=1\linewidth]{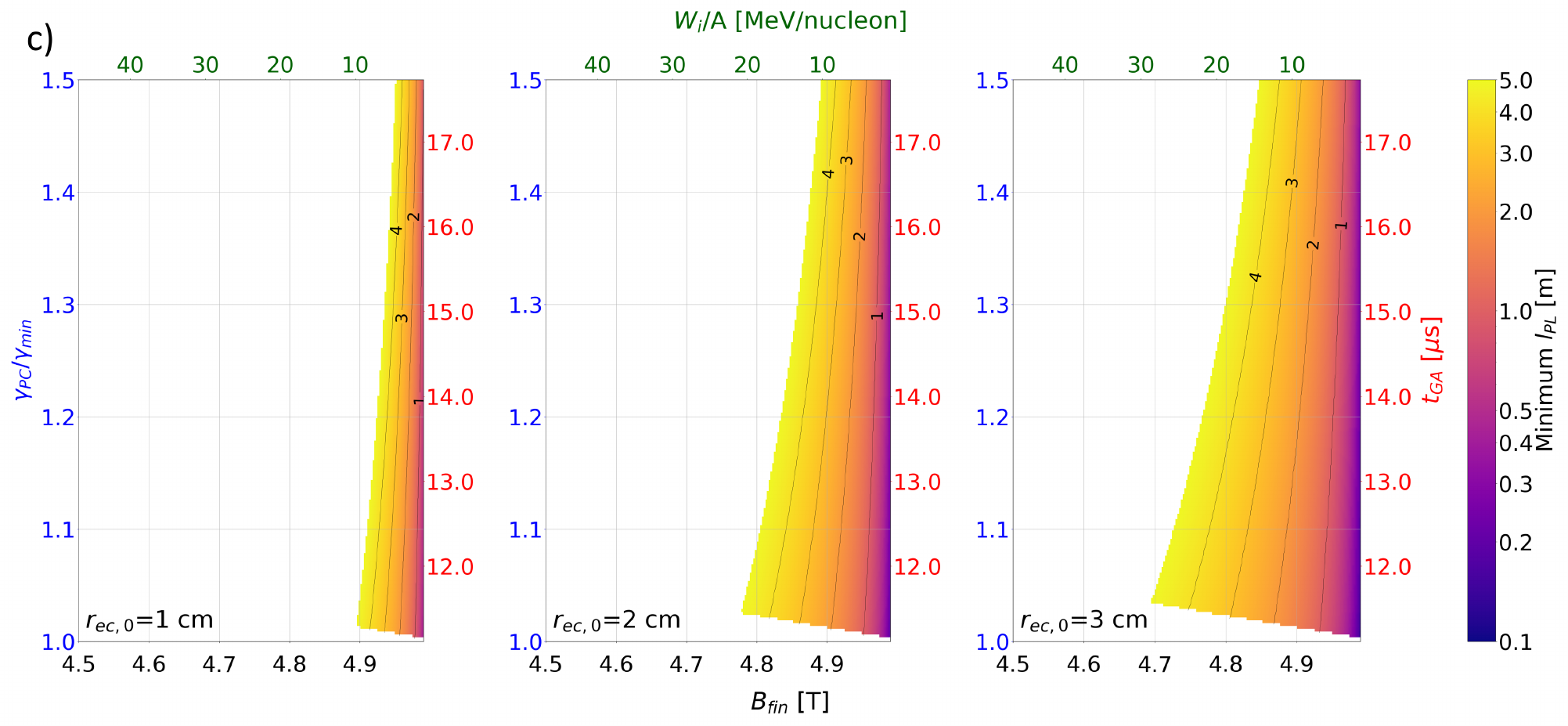}
\end{subfigure}
\caption{Stability maps of ECRIPAC for different parameters of the plasma, including a) the mass over charge ratio of the accelerated ion A/Z; b) the electron density in \% of critical density $n_{cr}$; c) the radius of the initial plasma disk $r_{ec,0}$. The nominal values of the accelerator parameters when they are not varied are $B_{max}=5$ T, $t_{pul}=50\;\mu s$, $f_{HF}=2.45$ GHz, A/Z = 2, $n_e=10 \% \; n_{cr}$ and $r_{ec,0}=2$ cm. The ion energy has been evaluated considering a negligible number of ions accelerated compared to the number of electrons (Eq.~\ref{eq:ecripac-ion_energy_simple}). The color bar representing the minimum cavity length ($l_{PL}$) required for a stable acceleration is in logarithmic scale.}
\label{fig:plasma_par}
\end{figure*}

This section investigates the set of admissible  physical parameter values for a stable ECRIPAC acceleration process. For convenience, the parameters are split into two groups: one for the applied external fields and one for the plasma.
The former includes the characteristics of the pulsed and PLEIADE fields, as well as the HF heating frequency. The latter focuses on the plasma density, dimensions, ion population (\textit{A/Z} ratio) and electron energy. 
\\
The influence of each parameter has been analyzed through a stability map (Fig.~\ref{fig:ext_fields} for the external fields parameters and Fig.~\ref{fig:plasma_par} for the plasma parameters), in order to find the minimum cavity length ($l_{PL}$) required for a stable acceleration, with $0.1 \leq l_{PL} \leq 5$ m. The latter is represented on the plots with a color bar in logarithmic scale.
In each map, the first y-axis (left side of each plot in blue) corresponds to the ratio between the normalized electron energy at the end of the plasma compression phase ($\gamma_{PC}$) and its limiting value $\gamma_{min}$ (Eq.~\ref{eq:ecripac_gamma_min}), with $1<\gamma_{PC}/\gamma_{min} < 1.5$.
The second y-axis (right side of each plot in red) is the time required to complete the GA phase ($t_{GA}$).
To evaluate $t_{GA}$, the necessary magnetic field at the end of the GA phase ($B_{GA}$) must be first estimated by combining Eqs.~\ref{eq:gyrac_autoresonance}, \ref{eq:ecripac_compression_gamma} and \ref{eq:ecripac_gamma_min} for the corresponding value of $\gamma_{PC}/\gamma_{min}$. Afterward, $t_{GA}$ can be obtained by inverting the equation for the time evolution of the magnetic field (in our study the following time evolution has been considered $B(t)=B_{max}\left[1-\frac{B_{max}-B_0}{B_{max}}\sin{\left(\omega_{pul}t+\frac{\pi}{2}\right)}\right]$, with $\omega_{pul}$ being the angular frequency of the pulsed coil).
The first x-axis (bottom side of each plot in black) indicates the final magnetic field at the end of the PLEIADE section $B_{fin}$, with $4.5\leq B_{fin}\leq 5$ T.
Finally, the second x-axis (top side of each plot in green) corresponds to the final values of the ion kinetic energy per nucleon $W_i/A$.
The nominal values of the accelerator parameters when they are not varied are $B_{max}=5$ T, $t_{pul}=50\; \mu$s, $f_{HF}=2.45$ GHz, A/Z = 2, $n_e=10 \% \; n_{cr}$ and $r_{ec,0}=2$ cm.
The ion energy in all stability maps has been evaluated using Eq.~\ref{eq:ecripac-ion_energy_simple} (considering $\gamma_{PC}>>1$), which implies a negligible electron energy variation and a much lower number of ions accelerated compared to the number of electrons.
In Fig.~\ref{fig:ext_fields}c, it can be observed that the upper part of the stability map for the case $f_{HF}=7$ GHz does not exist. Such a behavior is a consequence of the total rise time for the pulsed field $t_{pul}=t_{GA}+t_{PC}=const.$ (with $t_{GA}$ evaluated as explained above), considered constant in each study. Indeed, in the case of Fig.~\ref{fig:ext_fields}c, the time required by the GA phase $t_{GA}$ grows to a value larger than $t_{pul}$, and hence it is impossible to build an accelerator with such a design.

\subsection{External field parameters}

The external magnetic and electric fields of ECRIPAC are the main tools to control the accelerator operation. An overview of the effects of the maximum magnetic field value ($B_{max}$), the rise time of the pulsed field ($t_{pul}$) and the heating frequency ($f_{HF}$) can be observed in Fig.~\ref{fig:ext_fields}a, Fig.~\ref{fig:ext_fields}b and Fig.~\ref{fig:ext_fields}c respectively.
Increasing $B_{max}$ determines higher energies for the accelerated ions ($W_i/A$), shorter cavity length ($l_{PL}$) and a lower time required by the GA phase ($t_{GA}$) (see Fig.~\ref{fig:ext_fields}a).
For example, the maximum ion energy achievable for $l_{PL}=5$ m with the considered accelerator parameters is approximately 21, 22 and 23 MeV/nucleon for $B_{max}$ equal to 4.9, 5.0 and 5.1 T respectively. Since the effect of $B_{max}$ is not clearly visible in Fig.~\ref{fig:ext_fields}a, we reported a stability map for a much larger $B_{max}$ equal to 6 T in Appendix~\ref{app:pl}: in this case, the maximum achievable ion energy for the same cavity length and accelerator parameters becomes $\approx$29 MeV/nucleon, while an energy of 20 MeV/nucleon can be achieved with $l_{PL}=3.4$ m.
However, it also shifts the stability region for the accelerator towards larger values of the final magnetic field of the PLEIADE section ($B_{fin}$), adding technological constraints on the coil design.
On Fig.~\ref{fig:ext_fields}b, one can see that a lower rise time $t_{pul}$ has no significant effect on the accelerator stability, except reducing $t_{GA}$. A smaller $t_{pul}$ is therefore advantageous, as it enables a higher repetition rate in the accelerator’s pulsed-operation regime.
Nevertheless, the increase rate of the pulsed field over time has an upper limit, above which gyromagnetic autoresonance cannot take place~\cite{Golovanivsky-gyromagnetic_autoresonance}. Therefore, there exists a minimum value for $t_{pul}$ (which in turn determines the increase rate of the pulsed field) below which the GA phase cannot be achieved. Moreover, $t_{pul}$ is constrained by the technological limits of the pulsed coil: if $t_{pul}$ is too small, the resulting large Joule heating and hoop stress can severely damage the coil or even cause it to fail. To sum up, appropriate considerations must be made to choose a sufficiently large value for $t_{pul}$, taking into account both technological constraints of the pulsed coil and physical limitations of gyromagnetic autoresonance.
Finally, Fig.~\ref{fig:ext_fields}c plots show that a higher $f_{HF}$ during the GA phase has the benefit to lower the required $B_{fin}$, enlarging the stability region surface and reducing $l_{PL}$, allowing to accelerate ions at higher energies. This is a consequence of the larger plasma critical density ($n_{cr}=(\epsilon_0m_e(2\pi f_{HF})^2)/e^2$, where $\epsilon_0$ is the vacuum permittivity), leading to a higher number of electrons entraining the ions. On the other hand, it determines a much longer time for the GA phase due to the larger initial resonance field, which can overcome the maximum value imposed in this study for the duration of the first two phases (50 $\mu$s).
A general trend is observed for $B_{fin}$, used as one of the axes of the stability map: lower values provide larger energies for the ions at the cost of a longer PLEIADE section ($l_{PL}$). 


\begin{table*}[t!]
\centering
\caption{\label{tab:par_influence}%
Influence of each accelerator and plasma parameter on ECRIPAC operation, while the others are kept constant. A lower PLEIADE length $l_{PL}$ implies a larger stability region. WP stands for wave-plasma interaction.
}
\begin{tabular}{l c l l}
\hline\hline
Parameter & Parameter behavior & Advantages & Disadvantages \\
\hline
Maximum magnetic field ($B_{max}$) & $\uparrow$ &  $\uparrow$ $W_i/A$,  $\downarrow$ $t_{GA}$,  $\downarrow$ $l_{PL}$ &  $\uparrow$ $B_{fin}$ required \\
Rise time of the pulsed field ($t_{pul}$) & $\downarrow$ &  $\downarrow$ $t_{GA}$ &  Weaker slowly varying field hypothesis \\
Heating frequency of microwave ($f_{HF}$) & $\uparrow$ &  $\downarrow$ $l_{PL}$ & $\uparrow$ $t_{GA}$, Can $\downarrow$ Stability region \\
Final value of PLEIADE magnetic field ($B_{fin}$) & $\downarrow$ &  $\uparrow$ $W_i/A$ &  $\uparrow$ $l_{PL}$ \\
Mass over charge ratio (A/Z) & $\downarrow$ & $\downarrow$ $l_{PL}$, $\downarrow$ $t_{GA}$ & - \\
Electron density ($n_{e}$) & $\uparrow$ &  $\downarrow$ $l_{PL}$ &  WP problems near $n_{cr}$\\
Initial plasma disk radius ($r_{ec,0}$) & $\uparrow$ &  $\downarrow$ $l_{PL}$ &  $\uparrow$ $r_{cavity}$ \\
\hline\hline
\end{tabular}
\end{table*}

\subsection{Plasma parameters}

The effect of the plasma parameters on the accelerator stability is reported in Fig.~\ref{fig:plasma_par}.
It is clear that ECRIPAC is more suited to accelerate highly-charged ions with a relatively small mass (see Fig.~\ref{fig:plasma_par}a), since lower A/Z ratios lead to much larger stability regions and faster time for GA phase ($t_{GA}$).
Moreover, the acceleration is way more stable with higher plasma densities ($n_e$) and bigger dimensions of the initial plasma ($r_{ec,0}$) at a fixed density (see Fig.~\ref{fig:plasma_par}b and c respectively), due to the higher number of electrons and hence a better entrainment of the ions during PLEIADE. Nevertheless, particular attention must be devoted when considering these parameters: plasma densities close to the critical density can lead to problematic wave-plasma interaction, having a detrimental effect on the electron energy gain; larger dimensions for the initial plasma require instead the use of a bigger cavity for the GYRAC section of the accelerator (due to the larger $r_{ec}$ value during the GA phase), to avoid a fast deconfinement of the electrons nearby the plasma chamber walls.
Finally, the electron energy at the beginning of the PLEIADE phase ($\gamma_{PC}$) is more a requirement than a parameter for ECRIPAC, since it would be ideal to have the smallest value possible which allows stability. This allows faster times for the GA phase ($t_{GA}$) and shorter PLEIADE cavity ($l_{PL}$), minimizing the technological requirements for the accelerator and relaxing the conditions needed by the plasma.
A summary of the influence of each accelerator and plasma parameter is reported in Table~\ref{tab:par_influence}.


\section{Conclusion and prospects}
\label{sec:conclusion}

The theoretical study reported in the paper revisited and expanded the current knowledge of the ECRIPAC accelerator concept, highlighting further limitations for the accelerator stability and analyzing the effects that several parameters have on the overall acceleration process.
Errors in the calculations reported in the original paper~\cite{Geller-ecripac} have been addressed and cross-checked with ~\cite{Bertrand-report1,Bertrand-report2}. The correct results are presented in this work, providing a suitable reference for future investigations on this innovative accelerator concept.
\\
The minimum energy that the electrons must have at the end of the plasma compression phase in order to have a stable ion acceleration during the PLEIADE phase is actually higher than the previously reported limit (Eq.~\ref{eq:ecripac_gamma_min}), due to the increase in the plasma bunch radius and reduction in energy of the accelerated electron bunch. The existence of a stable acceleration process is also influenced by the maximum and minimum values of the PLEIADE magnetic field profile, a correct PLEIADE magnetic field profile ($\nabla B/B$) and the cavity length.
\\
A careful analysis of the effect that each parameter has on the acceleration process has been carried out, distinguishing external fields and plasma parameters. It has been noted that larger maximum magnetic fields and shorter rise time for the pulsed magnetic field have overall beneficial effects on the device, keeping in mind the technological and physical constraints related to these quantities. On the other hand, the heating frequency of the system must be carefully tuned, due to the presence of both positive and negative effects when increasing this parameter.
Furthermore, it has been concluded that ECRIPAC is more suitable to accelerate low A/Z ions, and that both a higher number of energetic electrons and a large electron density are extremely beneficial for the ion acceleration, despite possible problems related to wave-plasma interactions.
These results have been used in a complementary letter~\cite{Cernuschi-ecripac} to propose some practical designs able to accelerate several ions relevant for medical applications.
It must be emphasized that the current knowledge of the physics behind ECRIPAC considers mostly single particle dynamics, rarely taking into account the effects related to the presence of the full plasma. For this reason, an accurate numerical simulation is needed to properly assess and further expand the physics behind ECRIPAC.


\begin{acknowledgments}
We would like to express our sincere gratitude to Patrick Bertrand, a retired theoretician at the GANIL facility, for his valuable insights and extensive information regarding the early investigations of the ECRIPAC accelerator concept, which occurred during the nineties.
\end{acknowledgments}

\appendix


\section{Derivation of plasma compression phase formulas}
\label{app:comp}

The starting point to obtain the results presented in Section~\ref{sec:wp-comp} is the equation of motion of a charged particle under the influence of the Lorentz force, assuming some simplifications to simplify the treatment.
All symbols used in the various equations are summarized in Table~\ref{tab:app-symbols}.

\begin{table}[t]
\caption{\label{tab:app-symbols}
Notations for the various symbols used in the derivations.
}
\begin{ruledtabular}
\begin{tabular}{lr}
Quantity & Symbol\\
\colrule
Electric field & $\vec{E}$\\
Magnetic field & $\vec{B}$\\
Magnetic vector potential & $\vec{A}$\\
Particle position & $\vec{r}$\\
Particle velocity & $\vec{v}$\\
Particle momentum & $\vec{p}$\\
Normalized particle velocity & $\beta$\\
Particle orbit radius & r\\
Particle mass & m\\
Particle Lorentz factor & $\gamma$\\
Particle cyclotron angular frequency & $\Omega$\\
Elementary charge & e\\
Speed of light & c\\
\end{tabular}
\end{ruledtabular}
\end{table}

First, the magnetic field of the system is assumed to be directed exactly along the cavity axis (\textit{z}), quasi-uniform in space and slowly-varying in time (slowly varying field in time hypothesis), leading to an electric field induced by the time-varying magnetic field which can be expressed as (considering the Coulomb gauge-fixing condition)

\begin{equation}
    \vec{E}=-\frac{\partial \vec{A}}{\partial t}=\frac{1}{2}\vec{r}\times \frac{\partial \vec{B}}{\partial t} \; .
    \label{eq:app-Einduced}
\end{equation}

Second, due to the strong anisotropy of the electron velocity (resulting in the velocity parallel to the magnetic field being negligible with respect to the perpendicular component), the electron parallel motion is neglected and the equation of individual electrons, rotating exactly around the parallel z axis, is reduced to the transverse (\textit{x,y}) plane

\begin{equation}
    \begin{split}
    &\frac{d\vec{p}}{dt} = -e\left(E+v\times B\right) \\
    &= -e\left(\frac{1}{2}\vec{r}\times \frac{\partial \vec{B}}{\partial t} + \frac{\vec{p}}{\gamma m}\times\vec{B} \right) \;,
    \end{split}
    \label{eq:app-motionEq}
\end{equation}

where $\vec{r}$, $\vec{p}$ and $\vec{B}$ are all perpendicular vectors and the electric field is assumed to be given by Eq.~\ref{eq:app-Einduced}.
Since $\vec{r}\perp \vec{p}\perp \vec{B}$, the scalar multiplication of Eq~\ref{eq:app-motionEq} with $\vec{p}$ results in

\begin{equation}
    \frac{1}{2}\frac{dp^2}{dt}=-e\left(-\frac{1}{2}rp\frac{\partial B}{\partial t}\right) \;.
    \label{eq:app-motionEq_p}
\end{equation}

Remembering that

\begin{equation}
    \Omega =\frac{eB}{m\gamma} = \frac{p}{mr\gamma}\;,
    \label{eq:app-omega}
\end{equation}

it is possible to obtain starting from Eqs.~\ref{eq:app-motionEq_p} and \ref{eq:app-omega}:

\begin{eqnarray}
    \frac{dp^2}{p^2}=\frac{dB}{B} \;,
    \label{eq:app-pcost}
    \\
    r^2B=\frac{p^2}{e^2B} \;,
    \label{eq:app-rcost}
\end{eqnarray}

leading to Eq~\ref{eq:ecripac_compression_constants}.
By studying the motion along the x-axis and y-axis separately, it is possible to show that $r$ can be written in the general form

\begin{equation}
    r=r_{ec}+r_{Lar}e^{i\Omega t}
    \label{eq:app-r}
\end{equation}

and that the constant of motion is satisfied for both the component of the radius separately (Eq.~\ref{eq:ecripac_compression_constants}).
\\
From Eq.~\ref{eq:ecripac_compression_constants} it is straightforward to obtain Eqs.~\ref{eq:ecripac_compression_gamma} and \ref{eq:ecripac_compression_r}, considering $r=r_{Lar}+r_{ec}$ and remembering that \textit{p} can be written as $p=mc\sqrt{\gamma^2-1}$.
To obtain Eq.~\ref{eq:ecripac_compression_omega}, one must note that Eq~\ref{eq:app-omega} can be rewritten as $\Omega(t)=\frac{eB(t)}{m}\frac{1}{\gamma}$, which can easily lead to Eq.~\ref{eq:ecripac_compression_omega} by applying Eq.~\ref{eq:ecripac_compression_gamma}


\subsection{Dampening of axial oscillatory motion}

Let's approximate the magnetic mirror created by the pulsed magnetic field as a simple parabolic dependence along the \textit{z}-axis. Still considering the previous hypothesis, the total magnetic field can be written as $B_z(z,t)=B_{min}(1+\epsilon z^2+b(t))$, where $b(t)$ is a generic time dependence due to the pulsed magnetic field and $\epsilon=\frac{B_{max}-B_{min}}{B_{min}z_{mirr}^2}$ ($z_{mirr}$ is the extension of the magnetic mirror).
Since the magnetic field is divergenceless, its radial component can be written as

\begin{equation}
    B_r\approx -\frac{1}{2}r\frac{\partial B_z}{\partial z} \approx -B_{min}\epsilon rz\; .
    \label{eq:app-Br}
\end{equation}

Inserting Eq.~\ref{eq:app-Br} in the z-component of the equation of motion in cylindrical coordinates

\begin{equation}
    \frac{dp_z}{dt}=m\gamma\frac{d^2z}{dt^2}+m\frac{d\gamma}{dt}\frac{dz}{dt}=\frac{ep_{\theta}B_r}{m}\;,
\end{equation}

and rearranging the equation (considering an electron rotating around the cavity axis with $r_{Lar}=r=v_{\theta}/\omega_{HF}$ and remembering that during autoresonance $p_{\theta}^2\approx p^2=m^2c^2(\gamma^2-1)$)

\begin{equation}
    \begin{split}
    & \frac{d^2z}{dt^2}+\frac{\gamma}{\frac{d\gamma}{dt}}\frac{dz}{dt} = -\epsilon\frac{eB_{min}}{m}r\frac{m\gamma}{m\gamma}\frac{p_{\theta}}{m\gamma^2}z= \\
    & =-\epsilon\omega_{HF}r\frac{m\gamma}{m\gamma}\frac{p_{\theta}}{m\gamma^2}z=-\epsilon\frac{p_{\theta}^2}{m^2\gamma^3}z=\\
    & =-\frac{\epsilon m^2c^2(\gamma^2-1)}{m^2\gamma^3}z \;,
    \end{split}
\end{equation}

we obtain the equation of a dampened pseudo-harmonic oscillator for the axial position of the particle

\begin{equation}
    \frac{d^2z}{dt^2}+\frac{1}{\tau}\frac{dz}{dt}+\omega_{osc}^2z=0\;,
\end{equation}

where

\begin{equation}
    \tau = \frac{\gamma}{\frac{d\gamma}{dt}} \hspace{0.5cm} \omega_{osc}=c\sqrt{\epsilon\frac{\gamma^2-1}{\gamma^3}}\;.
\end{equation}

Therefore, the axial motion of the electrons during the plasma compression phase will be dampened in time.


\section{Derivation of PLEIADE phase formulas}
\label{app:pl}


\subsection{Constant of motion and ion energy}

Consider a single electron moving inside the PLEIADE magnetic field without any electric field. Using a cylindrical coordinate system $\hat{p}=p_r\hat{u}+p_{\theta}\hat{v}+p_z\vec{k}$ ($\hat{u}$, $\hat{v}$ and $\hat{k}$ are the unit vectors corresponding to the radial, azimuthal and axial direction respectively) the equation of motion can be rewritten separating the various components

\begin{eqnarray}
    \frac{dp_r}{dt}=p_{\theta}\left(\omega-\frac{eB_z}{\gamma m}\right) \; ,
    \label{eq:app-motionEq_rcomp}
    \\
    \frac{dp_{\theta}}{dt}=-p_r\left(\omega-\frac{eB_z}{\gamma m}\right)-\frac{ep_zB_r}{\gamma m} \; ,
    \label{eq:app-motionEq_thcomp}
    \\
    \frac{dp_z}{dt}=\frac{ep_{\theta}B_r}{\gamma m} \; ,
    \label{eq:app-motionEq_zcomp}
\end{eqnarray}

where $\omega=\frac{d\theta}{dt}$.
If the magnetic field gradient is sufficiently small ($B\approx B_z$), it is possible to consider the following adiabatic approximation

\begin{equation}
    \omega\approx\Omega = \frac{eB_z}{m\gamma} = \frac{p_{\theta}}{mr\gamma}\;,
    \label{eq:app-omega_pl}
\end{equation}

and the scalar multiplication of Eq.~\ref{eq:app-motionEq_thcomp} by $p_{\theta}$ can be rewritten as (using the first part of Eq.~\ref{eq:app-Br} and Eq.~\ref{eq:app-omega_pl})

\begin{equation}
    \frac{1}{2}\frac{dp_{\theta}^2}{dt}=-\frac{e}{\gamma m}p_{\theta}p_zB_r=\frac{1}{2}p_{\theta}^2\frac{\frac{\partial B_z}{\partial z}\frac{dz}{dt}}{B_z} \; .
    \label{eq:app-motionEq_p_pl}
\end{equation}

In analogy with Eqs.~\ref{eq:app-pcost} and \ref{eq:app-rcost}, Eqs.~\ref{eq:app-omega_pl} and \ref{eq:app-motionEq_p_pl} can be used to obtain the following constants of motion for the PLEIADE phase

\begin{equation}
    \frac{p_{\theta}^2}{B}=const \hspace{0.5 cm} r^2B=const \; .
    \label{eq:app-pl_const}
\end{equation}

Moreover, substituting Eq.~\ref{eq:app-motionEq_zcomp} in Eq.~\ref{eq:app-motionEq_p_pl} and integrating, it is possible to obtain the relation $p_z^2=p_{\theta,PC}^2-p_{\theta}^2$ (at the beginning of the PLEIADE phase $p_z\approx0$ and $p_{\theta}=p_{\theta,PC}$) which, after some manipulation and applying Eq.\ref{eq:app-pl_const}, leads to an evolution equation for the electron longitudinal velocity

\begin{equation}
    v_z = v_{\theta,PC}\sqrt{1-\frac{B_z(z(t),t)}{B_z(0,0)}} = v_{\theta,PC}\sqrt{1-\frac{B_z}{B_{max}}} \;.
    \label{eq:app-vz}
\end{equation}

Since during the PLEIADE phase the longitudinal velocities of electrons and ions have been observed to be approximately equal~\cite{Consoli-plasma_acceleration,Bardet-ion_entrainment}, Eq.~\ref{eq:app-vz} can be used to obtain Eq~\ref{eq:ecripac-ion_energy_simple} for an ion with mass $Am_a$ (considering a non-relativistic approach)

\begin{equation}
    \begin{split}
    &W_i \approx \frac{1}{2}(Am_a)v_z^2 \\
    &\approx \frac{1}{2}(Am_a)c^2\frac{\gamma_{PC}^2-1}{\gamma_{PC}^2}\left(1-\frac{B_z}{B_{max}}\right) \;,
    \end{split}
    \label{app:ion_energy}
\end{equation}

remembering that $v_{\theta,PC}^2=\frac{p_{\theta,PC}^2}{m^2\gamma_{PC}^2}\approx\frac{m^2c^2(\gamma^2_{PC}-1)}{m^2\gamma_{PC}^2}$.


\subsection{Non shake out condition}

An ion with mass $Am_a$ and charge $Ze$ can be accelerated by the electron population only if the ion acceleration due to Coulomb attraction is greater than the longitudinal acceleration of the electron bunch 
\begin{equation}
    \frac{(Ze)E}{Am_a}\geq \dot{v_z} \; .
    \label{eq:app-nso}
\end{equation}
By differentiating Eq.~\ref{eq:app-vz}

\begin{equation}
    \dot{v_z}=\frac{1}{2}v_{\theta,PC}^2\left| \frac{\nabla B_z}{B_{max}}\right|
\end{equation}

and inserting it in Eq.~\ref{eq:app-nso}, keeping in mind that $\frac{v_{\theta,PC}^2}{B_{max}}\approx\frac{v_{\theta}^2}{B_z}\leq \frac{c^2}{B_z}$ (due to Eq.~\ref{eq:app-pl_const} if the loss of electron energy is small), it is possible to obtain Eq.~\ref{eq:ecripac_nsc}.


\subsection{Electron bunch stability condition}

The stability of the electron bunch during the PLEIADE phase is determined by an inequality between the radial acceleration of the bunch ($\dot{v}_{rad}$) due to Coulomb repulsion and the axial acceleration of the bunch ($\dot{v}_{ax}$) along the cavity axis, where the two main contributions to the bunch stability are the relativistic dampening of Coulomb repulsion $F=\frac{eE}{\gamma^2}$~\cite{Reiser-eq_orbit} and the fast time needed by the electrons to axially cross the plasma chamber

\begin{equation}
    \begin{split}
    &\dot{v}_{rad}\leq\dot{v}_{ax}\; ,\\
    &\frac{1}{\gamma m}\frac{eE}{\gamma^2}\leq \frac{1}{2}v_{\theta,PC}^2\left| \frac{\nabla B_z}{B_{max}}\right| \;.
    \end{split}
    \label{eq:app-ebs}
\end{equation}

Considering as earlier $\frac{v_{\theta,PC}^2}{B_{max}}\approx\frac{v_{\theta}^2}{B_z}$ and $v_{\theta}\approx \beta c=c(1-\gamma^{-2})$, Eq.~\ref{eq:app-ebs} leads to Eq.~\ref{eq:ecripac_esc}.


\subsection{Progressive ion shake out}

Let's consider a plasma of a single element ($A$ is fixed) characterized by several ion charge-states, ranging from 1 to $Z_{max}$. $P(Z)$ is defined as the quantity of ions with charge $Z$ at the beginning of the PLEIADE phase.
It is now useful to rewrite the non shake out condition (Eq.~\ref{eq:ecripac_nsc}) highlighting the minimum charge-state $Z^*$ to satisfy the inequality, considering the maximum value of $\left|\frac{\nabla B_z}{B_z}\right|$ over the cavity length $z$

\begin{equation}
    Z^*(z)=\max_{0\leq\xi\leq z}\left(\frac{m_ac^2A}{2eE_{sc}}\left|\frac{\nabla B_z}{B_z}(\xi)\right|\right) \;.
\end{equation}

All ions with a charge-state $Z<Z^*(0)$ will not be accelerated during PLEIADE, and more ions will be lost during the temporal evolution of the PLEIADE phase depending on the value of $Z^*(z)$.
Hence, the energy balance of the system must be modified accordingly.
Rewriting the energy acquired by each ion (Eq.~\ref{eq:ecripac-ion_energy_simple}) at a considered cavity length $z$ as

\begin{equation}
    W_i(z)\approx\frac{1}{2}\frac{\gamma_{PC}^2-1}{\gamma_{PC}^2}Am_{a} c^2\left(1-\frac{B(z)}{B_{max}}\right) \;,
\end{equation}

the total energy gained by the ions $W_{i,tot}=\Delta W_{e\perp}-W_{e\parallel}$ will be repartioned between the ions which are shaken out of the plasma during the acceleration process and the ones which are accelerated until the end of the cavity.


\subsection{Estimation of the electron energy evolution during the PLEIADE phase}

Following Eq.~\ref{eq:energy_balance_pleiade}, the electron kinetic energy during the PLEIADE phase can be estimated considering only its perpendicular component ($W_e\approx W_{e,\perp}$), which in turn is determined by the electron azimuthal momentum $p_{\theta}$.
This approximation is quite realistic: considering an He$^{2+}$ accelerator design similar to the one considered in Section~\ref{sec:des_lim} ($\gamma_{PC}=16.69$) to accelerate ions up to 10 MeV/nucleon, the electron final axial velocity $v_z$ would be only $\approx$14\% of its initial azimuthal velocity $v_{\theta,PC}$ (Eqs.~\ref{eq:app-vz} and \ref{app:ion_energy}).
Hence, considering the constant of motion of the PLEIADE phase (Eq~\ref{eq:app-pl_const}), it is possible to estimate the reduction in the normalized electron energy $\gamma$ using the relativistic energy-momentum relation

\begin{equation}
    \begin{split}
    & \gamma=\sqrt{1+\frac{p^2}{m^2c^2}}\approx \sqrt{1+\frac{p_{\theta}^2}{m^2c^2}} \\
    & \approx\sqrt{1+\frac{p_{\theta,PC}^2}{m^2c^2}\frac{B_z}{B_{max}}}\;,
    \end{split}
\end{equation}

which in turn determines a more stringent condition for Eq.~\ref{eq:ecripac_esc} when considering the negative gradient magnetic field characterizing the PLEIADE section.


\subsection{Example of stability map with $B_{max}=6$ T}

\begin{figure}[H]
    \centering
    \includegraphics[width=1.\linewidth, valign=t]{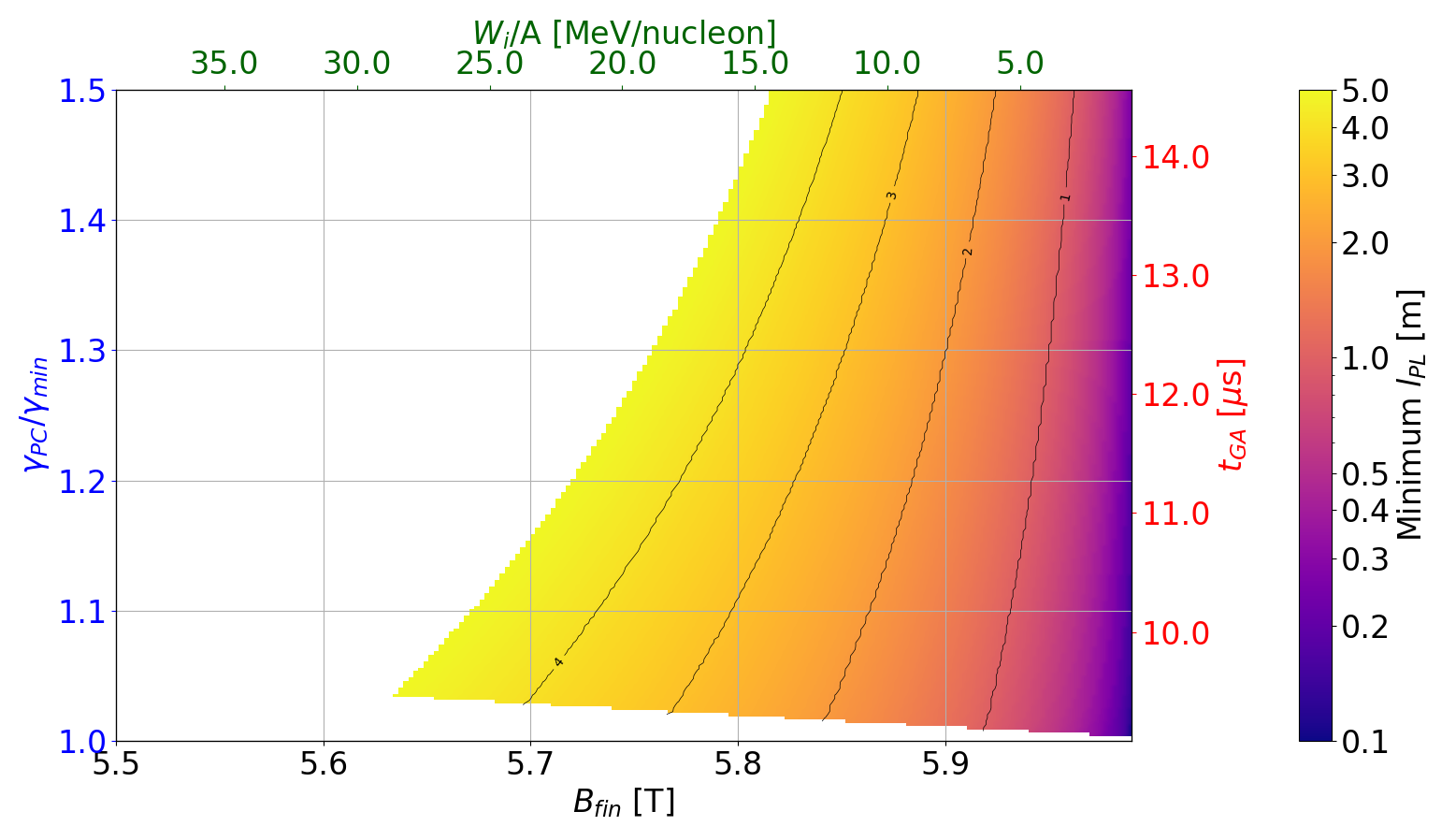}
    \caption{Stability map of ECRIPAC with $B_{max}=6$ T and all the other parameters set at their nominal value.}
    \label{fig:Bmax6T}
\end{figure}

Since the effect of the maximum magnetic field $B_{max}$ is not clearly visible in Fig.~\ref{fig:ext_fields}a, we reported in Fig.~\ref{fig:Bmax6T} an example of a stability map with $B_{max}=6$ T and all the other parameters set at their nominal value (see Sec.~\ref{sec:par_anal}), to better highlight the effect of $B_{max}$.

\providecommand{\noopsort}[1]{}\providecommand{\singleletter}[1]{#1}%

\clearpage
\end{document}